\documentclass[preprint]{aastex}

\shortauthors{Manset and Bastien}
\shorttitle{Periodic polarimetric observations of PMS binaries}

\begin{document}

\title{Polarimetric variations of binary stars. V. Pre-main-sequence
spectroscopic binaries located in Ophiuchus and Scorpius\altaffilmark{1}}  
\author{N. Manset\altaffilmark{2} and P. Bastien} \affil{D\'epartement
de Physique, Universit\'e de Montr\'eal,  C.P. 6128, Succursale
Centre-Ville, Montr\'eal, QC, H3C 3J7, Canada, and Observatoire du Mont
M\'egantic} \email{manset@cfht.hawaii.edu, bastien@astro.umontreal.ca}

\altaffiltext{1}{Based in part on observations collected with the 2m
Bernard-Lyot telescope (TBL) operated by INSU/CNRS and Pic-du-Midi
Observatory (CNRS USR 5026).  Financial support for the observations at
Pic-du-Midi was provided by the {\sl Programme National de Physique
Stellaire} (PNPS) of CNRS/INSU, France.}
\altaffiltext{2}{Now at: Canada-France-Hawaii Telescope Corporation,
65-1238 Mamalahoa Hwy, Kamuela, HI 96743, USA} 

\begin{abstract}
We present polarimetric observations of 7 pre-main-sequence (PMS)
spectroscopic binaries located in the $\rho$ Ophiuchus and Upper
Scorpius star forming regions (SFRs). The average observed polarizations
at 7660\AA\ are between 0.5\% and 3.5\%. After estimates of the
interstellar polarization are removed, all binaries have an {\it
intrinsic} polarization above 0.4\%, even though most of them do not
present other evidences for circumstellar dust. Two binaries,
NTTS~162814-2427 and NTTS~162819-2423S, present high levels of intrinsic
polarization between 1.5\% and 2.1\%, in agreement with the fact that
other observations (photometry, spectroscopy) indicate the presence of
circumstellar dust. Tests reveal that all 7 PMS binaries have a
statistically variable or possibly variable polarization. Combining these
results with our previous sample of binaries located in the Taurus,
Auriga and Orion SFRs, 68\% of the binaries have an intrinsic
polarization above 0.5\%, and 90\% of the binaries are polarimetrically
variable or possibly variable. NTTS~160814-1857, NTTS~162814-2427, and
NTTS~162819-2423S are clearly polarimetrically variable. The first two
also exhibit phase-locked variations over $\sim$10 and $\sim$40 orbits
respectively.  Statistically, NTTS~160905-1859 is possibly variable, but
it shows periodic variations not detected by the statistical tests;
those variations are not phased-locked and only present for short
intervals of time. The amplitudes of the variations reach a few tenths
of a percent, greater than for the previously studied PMS binaries
located in the Taurus, Orion, and Auriga SFRs. The high-eccentricity
system NTTS~162814-2427 shows single-periodic variations, in agreement
with our previous numerical simulations.  We compare the observations
with some of our numerical simulations, and also show that an analysis
of the periodic polarimetric variations with the Brown, McLean, \&
Emslie (1978) formalism to find the orbital inclination is for the
moment premature: non-periodic events introduce stochastic noise that
partially masks the periodic variations and prevents the BME formalism
from finding a reasonable estimate of the inclination.
\end{abstract}


\section{Introduction}
Pre-main-sequence (PMS) stars are objects still contracting to the main
sequence and surrounded by disks and/or envelopes of circumstellar dust
and gas. The dust grains produce polarization by scattering (see Bastien
1996 for a review). In PMS binary systems, the circumstellar matter may
be found around each star (circumstellar disk, or CS disk) and/or around
the binary itself (circumbinary disk, or CB disk).

We have shown (Manset \& Bastien 2000, 2001a, hereafter Papers~I and II)
how the linear polarization of a binary surrounded by circumstellar
matter varies periodically as a function of the orbital period, and how
the geometry of the disks, the nature and characteristics of the
scatterers, the masses of the stars, and the orbit characteristics
affect the polarimetric curves. Models can be used to find the orbital
inclination from those polarimetric variations (see for example Rudy \&
Kemp 1978; Brown, McLean, \& Emslie 1978). The work from Brown et
al. (hereafter BME) uses first- and second-order Fourier analysis
of the Stokes curves to give, in addition to the orbital inclination,
moments related to the distribution of the scatterers in the CS and CB
environments. The BME formalism was developed for Thomson scattering in
optically thin envelopes, and for binaries in circular orbits. Since
polarization in PMS stars is produced by scattering on dust grains, and
most of the known spectroscopic PMS binaries have eccentric orbits, the
BME formalism cannot be used a priori. However, we have shown (Papers~I
and II) that the BME analysis can still be applied in those cases, with
a few limitations.

In this context, we have obtained polarimetric observations of 24
spectroscopic PMS binaries. Detailed analyses were presented for the
only Herbig~Ae/Be binary of our sample, MWC~1080 (Manset \& Bastien
2001b, hereafter Paper~III), and for 14 PMS binaries located in the
Taurus, Orion, and Auriga SFRs (Manset \& Bastien 2002, hereafter
Paper~IV). MWC~1080's polarization and position angle are clearly
variable, at all wavelengths, and on time scales of hours, days, months,
and years. Stochastic variability is accompanied by periodic variations
caused by the orbital motion of the stars in their dusty
environment. The variations are not simply double-periodic (seen twice
per orbit) but include single-periodic (seen once per orbit) and
higher-order variations.

The analysis of 14 PMS binaries located in the Taurus, Auriga, and Orion
SFRs revealed that after removal of estimates of the interstellar
polarization, about half the binaries have an {\it intrinsic}
polarization above 0.5\%, even though most of them do not present other
evidences for the presence of circumstellar dust. Various tests reveal
that 77\% of those PMS binaries have or possibly have a variable
polarization. The polarimetric variations are noisier and of a
lesser amplitude ($\sim$0.1\%) than for other types of binaries, such as
hot stars. We have shown that an analysis of the periodic polarimetric
variations with the BME formalism to find the orbital inclination is for
the moment premature: non-periodic events introduce stochastic noise
that partially masks the periodic low-amplitude variations and prevents
the BME formalism from finding a reasonable estimate of the orbital
inclination.

Here we report the complete observations and detailed analysis for the
PMS binaries located in the $\rho$ Ophiuchus and Upper Scorpius SFRs.

\section{Observations}
The stars were chosen mainly from the list of binaries found in Mathieu
(1994), to which we added subsequent discoveries. Tables~\ref{Tab-Coord}
and \ref{Tab-Spectro} present basic information (other names,
coordinates, location), and spectroscopic and orbital data (spectral
type, PMS type, orbital period and eccentricity, orbital inclination
when known, and distance) for 7 of the binaries located in the $\rho$
Ophiuchus and Upper Scorpius SFRs.

The binaries were observed at the Observatoire du Mont M\'egantic (OMM),
Qu\'ebec, Canada, between 1995 May and 1999 June, using a $8\farcs2$
aperture hole and a broad red filter (RG645: 7660 \AA\ central
wavelength, 2410 \AA\ full width at half maximum). Polarimetric data
were taken with Beauty and The Beast, a two-channel photo-electric
polarimeter, which uses a Wollaston prism, a Pockels cell, and an
additional quarter-wave plate. The data were calibrated for instrumental
efficiency, instrumental polarization (due to the telescope's mirrors),
and zero point of position angle, using a Glan-Thomson prism,
non-polarized standard stars, and polarized standard stars,
respectively. On average, the instrumental polarization was very low
(less than 0.020\% $\pm$ 0.015\%, never above 0.03\%) and negligible,
which simplifies the determination of both instrumental polarization
and origin of position angle. The observational errors were calculated
from photon statistics, and also include uncertainties introduced by the
previously mentioned calibrations. The final uncertainty on individual
measurements of the polarization $P$ is usually in the range
0.03--0.05\%. The relative errors in position angle $\theta$ can be as
low as 0.1\arcdeg, but due to instrumental effects, systematic errors,
and the calibration procedure itself, the absolute errors on the
position angles are of the order of 1$^\circ$. For more details on the
instrument and the observational method, see Manset \& Bastien (1995,
2001b) and Manset (2000). Table~\ref{Tab-averpol} summarizes the
observations and gives the average polarization and position angle along
with the number of observations.

Data were also obtained in 1994 with the polarimeter STERENN at the
2-meter Bernard-Lyot Telescope of the Observatoire du Pic-du-Midi
(OPdM), France. This 2-channel polarimeter uses a half-wave plate
rotating at 20~Hz and a Wollaston prism, along with S-20
photo-multiplier tubes sensitive to the $UBV$ domain. A Johnson $V$
filter was used, along with a 10\arcsec\ aperture hole. Instrumental
efficiency and instrumental polarization were measured with polarized
and non-polarized standard stars, respectively. The observational errors
were calculated from the fit of the observations to a sinusoid function,
and range between 0.05\% and 0.10\% on individual measurements.

\section{Estimation of the interstellar polarization}
Polarimetric observations are usually a sum of interstellar (and
sometimes also intra-cluster) and intrinsic polarizations. An estimation
of the interstellar (IS) polarization for each object observed can be
used to assess the presence of intrinsic polarization. We have used the
Heiles (2000) catalog of over 9000 polarization measurements to
determine if the observed polarization for the PMS binaries studied here
is of intrinsic or IS origin, or a combination of both. This catalog is
an improvement over the one from Mathewson et al. (1978): it contains
additional observations, all data have been verified, and more precise
coordinates are given.

For each observed PMS binary, the catalog was scanned to select at least
$\approx$20 stars {\it with a similar distance}. Depending on the
stellar density and number of measurements in the catalog, this led to
the selection of a region between 6\arcdeg\ and 10\arcdeg\ in radius
around the target, and within 62 to 80~pc of it. The stars selected from
the catalog are used to compute an average of the IS polarization in
that region around the target, and also to find an average ratio of the
polarization to color excess $P/E(B-V)$. The IS position angle is
calculated with a simple average and with a distance-weighted average of
the polarization of all the stars selected; this method also gives more
weight to high IS polarization values (for which the position angle is
well determined) than to low polarization values (which have
poorly-determined position angles). In all the cases here, the two IS
position angle values are similar within their uncertainties, which
indicates that the alignment is generally good over all of the region
studied, and that the IS polarization value estimated is
reliable. However, if the IS position angles are not well aligned, the
average IS polarization value $P_{\rm IS}$ will be too low (since the
vectors cancel out). We therefore use a different method to find the IS
polarization\footnote{This is the reason why the errors for $P_{\rm IS}$
and $\theta_{\rm IS}$ are not related by their usual relation,
$\sigma_{\theta} = 26.85 \sigma_{P}/P$.}. Based on an extinction value
for the targets and assuming that this extinction is of IS origin only,
an estimate of the IS polarization is found from the average ratio $P/E(B-V)$.

If the position angles for the IS polarization and for the target are
different, it points to an intrinsic origin for at least part of the
polarization measured.  Intrinsic polarization is also deduced from
polarimetric variability.

Results are presented in Table~\ref{Tab-averpol}, where the weighted
averages of the observed $P$ and $\theta$ for the binaries are given
along with the possible origin of the observed polarization: a $\star$
symbol indicates intrinsic polarization while IS stands for interstellar
polarization\footnote{Table~\ref{Tab-averpol} includes the correct
interstellar polarization and intrinsic polarization estimates for the
binaries presented in Paper~IV, in which the values of $P_{\rm IS}$
given were computed with an inappropriate method (vectorial average of
the polarization). Thus, Paper~IV under-estimated the IS
polarization.}. When IS comes before a $\star$ symbol, the IS component
of the polarization is probably stronger than the intrinsic one, and
vice versa. The following columns present our calculation of the IS
polarization (polarization and position angle, along with their
uncertainties), based on the method presented above. We also give
$N_{IS}$, the number of measurements used to estimate this IS
polarization, and the radius and the interval of distance of the region
considered. Subtracting the IS polarization from the observed one gives
the intrinsic polarization, shown in the last columns.

Note that the IS polarization given in columns 6--9 of
Table~\ref{Tab-averpol} is only an {\it estimate} for the {\it whole
region} around a binary, and in some cases might not apply to a given
binary. In particular, it may include CS or intra-cluster material and
then over-estimates the IS polarization. Consequently, the intrinsic
polarizations given in the last columns should be considered crude
estimations only, intended to give an idea of the polarimetric
characteristics of the observed binaries as a whole, and not definitive
values of the intrinsic polarization for each binary.

To determine if an observed polarization has an IS component, we give
more weight to the value of $\theta_{\rm IS}$ deduced from neighboring
stars than to $P_{\rm IS}$ which depends on the value $E(B-V)$ given to
a target. Since $P_{\rm IS}$ can include the effect of CS material and
not just that of IS dust, it is less reliable. We do not have the
wavelength dependence of the polarization, which can usually be used to
extract the IS component. To help determine if there is an intrinsic
component of polarization, we also use the level of polarimetric
variability since IS polarization is stable.

\section{Polarimetric variability \label{p5-sec-var}}
Since a majority of single PMS stars are variable polarimetrically
(Bastien 1982; Drissen, Bastien, \& St-Louis 1989; M\'enard \& Bastien
1992), we also expected PMS binaries to be polarimetrically variable,
either periodically or not. We have found that 77\% of our 14 PMS
binaries located in the Taurus, Auriga, and Orion SFRs are variable or
suspected variable. Various tests were applied to check the polarimetric
variability or stability of PMS binaries: minimum and maximum values,
variance test, $Z$ test, and finally, a $\chi^2$ test.

One crude but easy way to check for variability in a set of observations
is to compare the difference between the maximum and minimum values of a
quantity with its average or typical observational uncertainty; variable
observations will have maximum and minimum values well outside the range
expected from the observational uncertainty. One can calculate the
variance of the sample (which is a measure of the ``width'' of the
observations, or of the scatter from the mean, or of the ``variability''
around a central value), and compare it to the standard deviation of the
mean, which gives the error from photon statistics as if all the
observations had been added together. For a set of observations of a
non-variable quantity, the sample variance will be low (the observations
are all clustered closely to the mean) and similar to the standard
deviation of the mean. But if there is variability, the ``width'' of the
observations will be greater than the standard deviation of the mean.

For the $Z$ test, if the data are ``well behaved'' or ``consistent'',
$Z\approx1$ within its standard error $\sigma_Z$. If $Z$ differs from
$1\pm\sigma_Z$, then there may be variability. Lastly, $\chi^2$ values
are calculated for $Q$ and $U$ separately, using 1$\sigma_i$ and
$1.5\sigma_i$. The probability of obtaining a given value of $\chi^2$ in
a Gaussian distribution is found for each of the four $\chi^2$
values. We define the criteria as follows: the star is variable if at
least 2 of the four $\chi^2$ values are over 0.95; the star is suspected
to be variable if one out of four $\chi^2$ values is over 0.95. We
developed in Paper~IV a protocol on how to interpret the results of
these tests in order to assign the polarization of a star as being
variable, suspected variable, possibly constant, or constant. We refer
the reader to Paper~IV for more information.

Many PMS binaries in this sample and in the Taurus, Auriga, Orion sample
(Paper~IV) show observations with polarization levels and/or position
angle well below or above the bulk of the data. We call these
observations ``atypical'' but they are nonetheless real. Close
examination of polarization observations taken over 5 years of
non-polarized standard stars (84 observations), polarized standard stars
(53 observations), 3 stars that were followed for many consecutive hours
(121 observations) did not show atypical observations like the ones we
repeatedly saw for PMS binaries (Manset 2000). Therefore, we believe
these atypical observations were due to some eruption-like events or
significant modifications in the CS environment (e.g.,
formation/destruction of condensations, accretion events), and not
because of instrumental problems. Since this study of binary PMS stars
is more concerned with orbit-induced, ``typical'', polarimetric
variations than with random, atypical ones, we have removed the atypical
observations before using the variability tests. This allows us to study
the more typical variations.

Table~\ref{Tab-AmpVar} presents the amplitude of the polarimetric
variations for all binaries of our sample, including MWC~1080
(Paper~III), the binaries located in the Taurus, Auriga, and Orion SFRs
(Paper~IV), and 2 systems which will be presented in future papers. The
amplitudes were calculated by simply taking the difference between the
minimum and maximum values of $P$, $\theta$, $Q$ and $U$; note that for
some cases, atypical observations were not considered so that the more
typical variations could be characterized. The variations (in $Q$ or
$U$) generally have amplitudes up to 0.3\%. Two binaries,
NTTS~162814-2427 and NTTS~162819-2423S, present much higher variations,
at the 1\% level. This indicates that, first of all, there is enough
dust in the environment of these 2 systems to produce the polarization,
and second, that the configuration is favorable to high-amplitude
variations. Since these variations are not clearly periodic, an
alternative explanation would rely on these systems being very active
(eruptive-like events, strong modifications of the CS environment) and
not on orbit-induced variations. In our previous numerical simulations,
the highest amplitude variations were produced with CS disks instead of
CB disks, but we were not able to produce variations above a few tenths
of a percent (Paper~II).

The details of the variability tests are shown in
Table~\ref{Tab-VarDetails}, where we give for each star the number of
observations used for the variability tests, $\sigma_{\rm sample}$ and
$\sigma_{\rm mean}$, $Z$ and its standard error, and $P{\chi^2}$,
calculated with $1\sigma$ and $1.5\sigma$. The conclusions of the
variability tests are shown in Table~\ref{Tab-Var}, where we have
classified the stars as ``variable'', ``suspected variable'', and
``possibly constant''. Of the 7 binaries, 3 are statistically variable
(NTTS~160814-1857, NTTS~162819-2423S, and NTTS~162814-2427) and the
others are suspected to be variable (NTTS~155808-2219, NTTS~160905-1859,
Haro~1-14C, NTTS~155913-2233). When combined with the results from
Papers~III and IV, 90\% of the PMS binaries (19/21) are variables or
suspected variables, in agreement with the results found by Bastien
(1988) and M\'enard \& Bastien (1992) for a sample of mostly single PMS
stars.

\section{Periodic polarimetric variations \label{p5-sec-pervar-gen}}
In addition to the general variability, which is a known property of
single PMS stars, PMS binaries will also present periodic polarimetric
variations caused by the orbital motion, even if in some cases the
amplitude may be too small to be detected with the currently available
instruments or masked by non-periodic or pseudo-periodic variations. In
the case of Mie scattering, we have also shown in Paper~II that dust
grains, which are mostly responsible for the polarization, are less
efficient polarizers and produce smaller amplitude variations than
electrons. This is an indication that periodic polarimetric variations
could be more difficult to observe in PMS binaries than in, for example,
hot binaries surrounded by electrons, which can easily exhibit
variations of a few tenths of a percent (see for example Robert et
al. 1990; Robert et al.  1992). The size of the grains also determines
the amplitude of the polarimetric variations: dust grains with radii
$\sim 0.1 \micron$ produce the largest polarimetric variations
(Paper~II).

To look for periodic polarimetric variations, the known orbital periods
are used to calculate the orbital phase for each measurement. The
polarization $P$, its position angle $\theta$, and the Stokes parameters
$Q$ and $U$ are plotted as functions of the orbital phase (see
Figures~\ref{Fig-n155808}, \ref{Fig-n155913a} $-$ \ref{Fig-n160814},
\ref{Fig-n160905a} $-$ \ref{Fig-haro}, \ref{Fig-n162814a} $-$
\ref{Fig-n162819}). When enough data are 
available, observations are represented as first and second harmonics of
$\lambda=2\pi\phi$, where $\phi$ is the orbital phase:\\
\begin{eqnarray}
Q &=& q_0 + q_1 \cos \lambda + q_2 \sin \lambda + q_3 \cos 2\lambda +
q_4 \sin 2\lambda, \label{p5-eq-qfit}\\
U &=& u_0 + u_1 \cos \lambda + u_2 \sin \lambda + u_3 \cos 2\lambda +
u_4 \sin 2\lambda. \label{p5-eq-ufit}
\end{eqnarray}
The coefficients of this fit are then used to find the orbital
inclination, following the BME formalism and using the first and second
order Fourier coefficients:\\
\begin{eqnarray}
\left[ \frac{1-\cos i}{1+\cos i} \right]^2 &=& \frac{(u_1+q_2)^2 +
(u_2-q_1)^2}{(u_2+q_1)^2 + (u_1-q_2)^2} \label{EQ-iO1-p5} , \\
\left[ \frac{1-\cos i}{1+\cos i} \right]^4 &=& \frac{(u_3+q_4)^2 +
(u_4-q_3)^2}{(u_4+q_3)^2 + (u_3-q_4)^2} \label{EQ-iO2-p5}.
\end{eqnarray}
The representation with sinusoids is a useful representation
of orbit-induced variations, and is not used only to
apply the BME formalism.


For 5 of the 7 binaries, we have enough data to look for periodic
variations , and some of them do show periodic variations (see
below). However the curves representing the fits made according to
Eq.~\ref{p5-eq-qfit} and \ref{p5-eq-ufit} do not always represent the
observations well, for various reasons. Effects induced by dust instead
of electrons as scatterers, elliptical instead of circular orbits, and
variable absorption effects may produce variations which may not be well
represented by the first and second harmonics only. The fitted curves
are nevertheless a useful representation of the data and illustrate how
the BME formalism, extended by the numerical simulations of Papers~I and
II, can be applied.

Periodic polarimetric variations can be caused by the binarity (orbital
motion) or the presence of hot or cool stellar spots, among a few
reasons. Classical T~Tauri Stars (CTTS) are known to have both cool and
hot spots, and WTTS generally have only cool spots (Bouvier et
al. 1993), some of which can be stable over periods of several months
(on V410 Tau for example, Herbst 1989). Since all the binaries observed
here are WTTS and in general display only small photometric variations,
we believe that the spots causing the photometric variations, if
present, are small, and have a very small effect on the polarization. In
addition to stellar spots, non-periodic phenomenons such as eruptive
events, variable accretion and rearrangements of the CS or CB material
can cause pseudo-periodic polarimetric variations that may mask the
strictly periodic ones, especially if the observations are taken over
many orbital periods, as is the case here.

Despite these difficulties, some binaries present periodic variations:
NTTS~160814-1857, \linebreak NTTS~162814-2427, NTTS~160905-1859 (for short
intervals of time), and NTTS~155913-2233 (low-amplitude variations seen when
the data are binned). To investigate the significance of this
periodicity, a Phase Dispersion Method (Stellingwerf 1978) and a Lomb
normalized periodogram algorithm (Press et al. 1997) were used. The
Phase Dispersion Method (PDM) is a least-squares fitting technique
suited for non-sinusoidal time variations covered by irregularly spaced
observations, and finds the period that produces the least scatter about
the mean curve. The Lomb normalized periodogram (LNP) method is more
powerful than Fast Fourier Transform methods for uneven sampling, but
still assumes the curve is sinusoidal, which may not be always
appropriate for the polarimetric observations presented here. The
periods found by using both methods are very similar to one another for
a given star, but the significance is usually marginal.

NTTS~162814-2427 presents single-periodic variations readily seen in
position angle and $U$ (see Figure~\ref{Fig-n162814b}). The PDM finds in
the position angle data a period of $35.4d$, and the LNP has two peaks,
at 35.7 and $32.7d$, but those 3 peaks are local extrema with other
periods coming up with similar significance. Therefore, we cannot
attribute a strong significance to those peaks. When using the binned data
instead of the whole set of observations, the LNP recognizes the
presence of periodicity and the period found has a 60\% probability of
not coming from random Gaussian noise, which is only slightly better
than chance. Those periods compare rather well with the known orbital
period of $35.95d$.  NTTS~162814-2427 does show variations with
amplitudes of a few tenths of a percent. Our numerical simulations can
produce amplitudes of $\sim$0.1\% if there is a CS disk but no CB disk.
NTTS~162814-2427 shows strong single-periodic variations, which we
attribute to the large orbital eccentricity. For most of the other
binaries, the presence or absence of $1\lambda$ variations, and the
exact cause of the single-periodic variations, if present, are harder to
determine.

We will now discuss individual stars.

\section{Comments on individual stars \label{p5-sec-stars}}
The detailed observations are presented in Tables~\ref{Tab-n155808} to
\ref{Tab-n162819} and in Figures~\ref{Fig-n155808} to \ref{Fig-n162819}.


\subsection{NTTS~155808-2219 = ScoPMS~20}
This 16.95-day binary (Mathieu 1994) has a projected separation of
0.048~AU (Jensen, Mathieu, \& Fuller 1996), and there is a tertiary
component (Mathieu, Walter, \& Myers 1989; Walter et al.  1994). The
radial velocity measurements exclude membership in the Upper Scorpius
association (Walter et al.  1994). The primary has a rotation period of
4.30 d (Adams, Walter, \& Wolk 1998).

Data for NTTS~155808-2219 are presented in Table~\ref{Tab-n155808} and
Figure~\ref{Fig-n155808}. The statistical tests conclude that the
average polarization at 7660\AA, 0.48\% at 139\arcdeg, is possibly
variable (see Tables~\ref{Tab-VarDetails} and \ref{Tab-Var}). This
conclusion is based on only four observations that show a constant
polarization but one position angle well above the others. From the
polarization catalog of Heiles (2000), we find an average IS
polarization of $0.34\pm0.07$\% at 88\arcdeg\ (see
Table~\ref{Tab-averpol} and the map on Figure~\ref{Fig-sco3}). This
value was found by averaging the polarization of 22 neighboring stars
located within 6\arcdeg\ and 80~pc of NTTS~155808-2219. The binary
presented in the next section, NTTS~155913-2233, is 0.35\arcdeg\ from
NTTS~155808-2219 and also has a very similar polarization, 0.49\% at
124\arcdeg, which points to a common IS polarization for both
binaries. Since the observed and IS position angles are different and
NTTS~155808-2219 is possibly variable, we conclude that its polarization
is the sum of intrinsic and IS polarization.

\subsection{NTTS~155913-2233 = ScoPMS~23} 
This 2.42378-day binary (Mathieu et al. 1989) is a double-lined system
(Prato \& Simon 1996) with a separation of 0.014~AU (Jensen et
al. 1996). A weak tertiary component was detected spectroscopically
(Mathieu et al. 1989) and found to be at large distance from the binary
since its effect on the velocities of the binary or the tertiary itself
was not detected. It was later detected in a 2.2 \micron\ speckle
imaging survey, at a separation of 0\farcs288 (or 45~AU at a distance of
160~pc) and position angle 347\arcdeg\ (Ghez, Neugebauer, \& Matthews
1993). It is included in our $8\farcs2$ aperture hole. The two stars of
the spectroscopic binary are K5 and M5 stars, the third star is a K5,
and the mass ratio of the spectroscopic binary is 2.1 $\pm$ 0.24 (Prato
\& Simon 1996). The primary has a rotation period of 3.30 d (Adams et
al. 1998). There is no evidence for extended circumstellar material,
either in low-dispersion spectroscopic or IR photometric data; there is
no evidence for an associated disk, mass accretion or mass loss (Mathieu
et al. 1989). Although it is young, with an age of $10^6$ yr (Mathieu et
al. 1989), it is a mature system, with a circular orbit and no evidence
for circumstellar material.

Data are presented in Tables~\ref{Tab-n155913} and \ref{Tab-n155913-v},
and in Figures~\ref{Fig-n155913a}, \ref{Fig-n155913b}, and
\ref{Fig-pic155913}. The polarization is similar at 7660\AA\ and
5550\AA. Statistical tests do not indicate any variability at 5550\AA,
but this might be due to the small number of observations and the
relatively high uncertainties. At 7660\AA, the binary seems to be more
variable than at 5550\AA; statistical tests indicate that
NTTS~155913-2233 is possibly variable
(Table~\ref{Tab-Var}). Figure~\ref{Fig-n155913a} shows variations with
an peak-to-peak amplitude of 0.2\% in $P$, to be compared with the
average uncertainty 0.042\%. No periodic variations are seen. There
might be two different epochs, one from 1995 May to 1996 May and having
a peak in polarization at phases 0.3-0.4, and the second from 1997 April
to 1997 June, with a peak at phases 0.5-0.6. However, the data from
these two epochs do not show any more periodic variations than all of
the data combined. Figure~\ref{Fig-n155913b} presents the observations,
where the period has been divided into 20 bins, and the data averaged in
each bin. The binned data may show low-amplitude periodic
variations. The orbit is almost circular, but variations are not purely
double-periodic. Data obtained at 5550\AA, presented in
Figure~\ref{Fig-pic155913}, show a systematic (as opposed to random)
variation of the position angle. These variations cover a different time
interval and are not similar to those seen at 7660\AA.

Two atypical observations are not shown in the figures for the 7660\AA\
data and were not used in the variability calculations. These 2
measurements were taken almost 2 years apart, but have similar position
angles, 162\arcdeg\ and 168\arcdeg, very different from the rest of the
observations (124\arcdeg). The first atypical observation has a
polarization lower than the rest of the data, but the second has a
polarization very similar to the general polarization observed. We
believe these 2 atypical observations are evidence for significant
transient changes in this system, an eruptive-like event at the surface
of one of the stars or a change in the matter distribution, for
example. Such atypical polarizations or positions angles have been
observed before in the HAeBe binary MWC~1080 (Paper~III) and other PMS
binaries (Paper~IV). These non-periodic variations are also in agreement
with the general polarimetric variability observed in single PMS. Since
NTTS~155913-2233 is 0.35\arcdeg\ from NTTS~155808-2219 (see map in
Figure~\ref{Fig-sco3}), has a very similar polarization, and is possibly
constant, we conclude that the observed polarization is the sum of
intrinsic and IS polarization.

\subsection{NTTS~160814-1857 = HBC~630 = ScoPMS~44}
Although this 144.7-day binary (Mathieu 1994) is identified as 'tt' in
the HBC catalog, Walter (1986) identifies it as a NTTS; its H$\alpha$
line is in emission, with $W_{\lambda}=0.6$ \AA. The projected
separation of this single-lined spectroscopic binary is 0.19~AU (Jensen
et al.  1996). It is not found in the IRAS Point Source Catalog but
there is a flat NIR excess that can be modeled by adding to the K2~IV
primary a cool M3 IV companion 3.7 mag fainter than the primary at $V$,
although this is not a unique solution; the non spectroscopic detection
of the secondary suggests that the luminosity ratio exceeds 10:1 (Walter
et al.  1994). According to Zakirov et al.  (1993), brightness
variations in $V$ had an amplitude of 0.16 mag in 1991/1992, with a
rotational period of 3.81 d; no eclipsing effect could be
found. Spectroscopic evidence argues against the presence of significant
circumstellar material (Walter 1986).

Data are presented in Table~\ref{Tab-n160814} and
Figure~\ref{Fig-n160814}. The average polarization at 7660\AA, 1.94\% at
117\arcdeg, is statistically variable (see Table~\ref{Tab-VarDetails}
and \ref{Tab-Var}); the single observation made at 5550\AA\ is lower
than the 7660\AA\ values, by $\sim 0.3$\%, but the position angle is
similar. Even though spectroscopic observations do not indicate the
presence of CS material (Walter 1986), the intrinsic polarization is
significant (0.7\%) and variations are present. The nine observations taken at
7660\AA, which were taken over $\approx$ 8 orbits, are very well fitted
by the theoretical sinusoidal curves, and do not show the scatter
usually observed. For example, GW~Ori, for which we have 11 observations
taken over $\approx$ 4 orbits, does not show variations as well behaved
as those in NTTS~160814-1857 (Paper~IV). Differences in the CS and CB
environment (configuration or stability) are probably the
explanation. The catalog from Heiles (2000) shows that the IS
polarization is lower than the observed one, $1.51\pm0.26$\% at
123\arcdeg\ (see map in Figure~\ref{Fig-sco4}). The similar IS position
angle indicates the presence of IS polarization. Since NTTS~160814-1857
is polarimetrically variable, we conclude that its polarization is the
sum of strong IS and weaker intrinsic components.

\subsection{NTTS~160905-1859 = HBC~633 = ScoPMS~48}
This 10.4-day binary (Mathieu 1994) is also identified as 'tt' in the
HBC catalog but Walter (1986) identifies is as a NTTS; the H$\alpha$
line is seen in {\it absorption} with $W_{\lambda}=0.6$ \AA. The
projected separation of this single-lined spectroscopic binary is
0.015~AU (Jensen et al.  1996). It is not found in the IRAS Point Source
Catalog but the small excess in NIR (smaller than the one for
NTTS~160814-1857), could be due to the companion, which would then be 4
mag fainter at $V$, although this is not a unique solution.
Non-detection of the secondary in spectroscopic observations suggests
that the luminosity ratio exceeds 10:1 (Walter et al. 
1994). Spectroscopic evidence argues against the presence of significant
circumstellar material (Walter 1986). The derived age, $7 \times 10^6$
yr, makes it one of the oldest WTTS (Mathieu et al. 1989).

NTTS~160905-1859 is 0.2\arcdeg\ from NTTS~160814-1857 (see map in
Figure~\ref{Fig-sco4}). It has a slightly lower polarization average of
1.38\% at 134\arcdeg\ (see Table~\ref{Tab-averpol}). It is a suspected
variable (see Table~\ref{Tab-Var}). As for the previous binary, we
conclude that its polarization is the sum of IS and intrinsic
components. 

Data are presented in Tables~\ref{Tab-n160905} and \ref{Tab-n160905-v},
and in Figure~\ref{Fig-n160905a} to \ref{Fig-n160905e}. As was the case
for NTTS~160814-2427, the polarization at 5550\AA\ is lower than the
observations taken at 7660\AA\, by about $\sim 0.3$\%, but the position
angles are all very similar. This binary shows some signs of variability
at 7660\AA\, but not at 5550\AA. The first figure shows all the data
taken at 7660\AA\ except two atypical observations that stand out from
the bulk of the data. These two measurements, taken 2 years apart,
present position angles (141\arcdeg\ and 121\arcdeg) different than the
average (134\arcdeg) although the polarization levels are typical of
this star. These atypical observations are the result of some
non-periodic change in the CS or CB environment. The figure also shows
that there is a lot of scatter and no obvious periodic variations. If we
divide the data in different epochs of observations, we see periodic (or
at least regular) behavior for a given epoch, but the behavior changes
from one epoch to the other. Observations taken between 1997 April 2 and
April 16 show smooth variations (see Figure~\ref{Fig-n160905b}). During
that period, the polarization was maximum at phase 0.1 and 0.7, whereas
during the period from 1997 June 3 to July 11 (see
Figure~\ref{Fig-n160905c}), a maximum in polarization was seen at phase
0.3 and a minimum at phase 0.0; the period we used, $10.400 \pm 0.002$ d
(Mathieu et al. 1989), is too well determined to allow shifts in phase,
so a rearrangement must have occurred between the two sets of data,
taken only 2 months apart. Figure~\ref{Fig-n160905e} is the same as
Figure~\ref{Fig-n160905d} except that it includes data taken between
1998 April 27 and May 13; the added observations are all below the
others. In this last figure, variations may not be as smooth as in
Figure~\ref{Fig-n160905b}; variations in position angles are almost
single-periodic with the exception of one point near phase 0.45. These
sets of data from different epochs show different average polarization
values and different polarization variations, which might indicate
changes in the CS/CB environment for this star. This makes the
accumulation of data over many years very noisy (compare the data for
each epoch with Figure~\ref{Fig-n160905a}).

\subsection{Haro 1-14C = HBC~644}
This 591-day binary has an average polarization of 1.08\% at 34\arcdeg
(Table~\ref{Tab-averpol}). Data are presented in Table~\ref{Tab-haro}
and Figure~\ref{Fig-haro}, where one atypical observation is not
shown. It is a suspected variable (Table~\ref{Tab-VarDetails} and
Table~\ref{Tab-Var}).  The average IS position angle of stars with
the same distance modulus is 24\arcdeg, similar to the observed position
angle and indicative of IS polarization (see map in
Figure~\ref{Fig-sco5}). Since this star may be variable, we conclude
that its polarization has a strong IS and a weaker intrinsic components.

\subsection{NTTS~162814-2427 = ROX~42 or ROX~42C}
NTTS~162814-2427 is a triple system, with a companion discovered by Lee
(1992), at 0\farcs15 and position angle 135\arcdeg\ (Ghez et al.  1993)
and thus included in our measurements. The projected separation of the
35.95-day spectroscopic binary (Mathieu 1994) that has a mass ratio of
1.09$\pm$0.07 (Lee, Mart\'{\i}n, \& Mathieu 1994) is 0.27~AU, while the
third star is at 19.4~AU (Jensen et al.  1996). This star has one of the
most eccentric orbits. It also has a small UV excess, a small NIR excess
and a weak H$\alpha$ line with a P Cygni profile (Walter et al.  1994),
which might indicate that there is more CS material than for other WTTS,
but not enough to classify the star as CTTS. The shape of the H$\alpha$
emission line changes with time, sometimes showing inverse P~Cygni
profile, which might be an indication of accretion processes (Mathieu
1992) with a low accretion rate.  The age derived from the position in the
HR diagram gives $1\times10^6$ yr, which makes this binary one of the
youngest yet found (Mathieu et al. 1989). Theoretical
evolutionary models with different input physics have been investigated
by Figueiredo (1997) who found that the observations for this binary are
well fitted by a 1.10 M$_{\sun}$ K4 primary, and a K5 1.00 M$_{\sun}$
secondary, both with an age of $3.7 \times10^6$ yr. NTTS~162814-2427 was
not detected at 1100 \micron\ (Skinner, Brown \& Walter 1991), so an
upper limit to the disk mass based on the $3\sigma$ upper value of the
1100 \micron\ observations is 0.11 M$_{\sun}$, although this star was
observed under poor atmospheric conditions. There is a significant
excess from the NIR up to 60 \micron, consistent with the presence of CB
material (Walter et al. 1994)

Its SED is consistent with a CB disk in which the central region has
been cleared (Jensen \& Mathieu 1997). Using theoretical masses derived
from evolutionary tracks and spectroscopic observations, Jensen \&
Mathieu (1997) find and use an inclination of 71\arcdeg. The semi-major
axis is then 0.28~AU, and the dynamically cleared gap goes from 0.057 to
0.85~AU. No silicate feature appears near 10 \micron. The SED modeling
shows that there is no requirement that the disk be optically thin. The
dynamical gap fits the data better than a continuous disk. The best fit
gap is similar to the dynamical gap predicted by theory; the best fit is
for 0.047~AU to 0.40~AU. The dynamical hole does not reproduce the
observations well. But the model does not take into account the presence
of the third star. The IR observations could be due to emission coming
from the tertiary or its CS disk, without the need for disks around the
primary and/or secondary stars. Also, the lack of submm emission could
come from a cleared up CB disk, caused by interaction with the
tertiary, and not from a gap in the CB disk.

The $V$ magnitude varies up to 0.4 mag (Lee 1992 - reported by Jensen
\& Mathieu 1997). According to Zakirov et al.  (1993), brightness
variations in $V$ had an amplitude of 0.32 mag in 1991/1992, with a
rotational period of 9.32d; no eclipsing effect could be found. The
similarity of the dynamical mass limits and the theoretical masses
suggests that the inclination is small (Mathieu et al. 1989),
but, as mentioned above, Jensen \& Mathieu (1997) found an inclination
of 71\arcdeg.

NTTS~162814-2427 is at 0.5\arcdeg\ from Haro~1-14C (see map in
Figure~\ref{Fig-sco5}). Its high polarization (3.5\% at 22\arcdeg) and
high variability (see Tables~\ref{Tab-AmpVar}, \ref{Tab-VarDetails} and
\ref{Tab-Var}) indicate a large intrinsic polarization. The average IS
position angle of stars with a similar distance modulus is 25\arcdeg,
close to NTTS~162814-2427's value and indicative of IS
polarization. NTTS~162819-2423S has a polarization position angle close
to that for NTTS~162814-2427, so this might indicate a local
(intra-cluster) origin for part of the polarization of these two
stars. We conclude that NTTS~162814-2427 has strong intrinsic
polarization with a weaker IS component and possibly with a strong
intra-cluster polarization.

Data are presented in Table~\ref{Tab-n162814} and
Figures~\ref{Fig-n162814a} and \ref{Fig-n162814b}. The polarization has
an average of 3.5\% at 22\arcdeg and is statistically variable. Two
atypical observations are not included in the figures. The first one has
a polarization almost 1\% below the average, and a position angle of
15\arcdeg. The second atypical observation also has a position angle of
15\arcdeg, but its polarization level is only 0.3\% below the
average. These atypical measurements indicate some non-periodic change
in this system.  Subtraction of the IS component gives an intrinsic
polarization of 2.1\%, rather high for a WTTS or any PMS star. The
presence of intrinsic polarization is in agreement with the evidences
for CS material, but part of the intrinsic value could still be intra-cluster
polarization.  The fact that the amplitude of the polarimetric
variations is very large, up to 0.5\% (see Figure~\ref{Fig-n162814a} and
\ref{Fig-n162814b}), which is higher than for the other WTTS we
observed, may also be supporting the indication that there is more CS
material around NTTS~162814-2427 than for other WTTS. Alternatively, the
high polarimetric variations may be due to the high eccentricity, or a
favorable inclination. The binned data clearly show a periodic variation
in position angle. In position angle and in $U$, the observations are
strongly dominated by single-periodic variations; this might be a direct
consequence of the orbital eccentricity, which is equal to 0.48. Strong
single-periodic variations are here a consequence of orbital
eccentricity, since the mass ratio is near unity and we have shown that
in these cases, asymmetric CB envelopes do not produce single-periodic
variations (Paper~I).

\subsection{NTTS~162819-2423S = ROX~43A}
NTTS~162819-2423 is a quadruple system, with both components
NTTS~162819-2423S (ROX~43A) and NTTS~162819-2423N (ROX~43B) themselves
binary stars.  The projected separation of the 89.1-day spectroscopic
binary (Mathieu 1994) (south component) is 0.10~AU, whereas the binary
in the north component has a projected separation of 2~AU and both
binary systems are separated by 600~AU (projected) (Jensen et al.
1996). NTTS~162819-2423N is about 1~mag fainter in $V$ than the southern
component (Mathieu et al. 1989) and $4\farcs8$ away (Simon, Ghez, \&
Leinert 1993). According to Zakirov et al.  (1993), brightness
variations in $V$ had a small amplitude of 0.1 mag in 1991/1992, with a
rotational period of roughly 3.2 d; no eclipsing effect could be found.

NTTS~162819-2423S was not detected at 1100 \micron\ (Skinner et
al. 1991), so an upper limit to the disk mass based on the $3\sigma$
upper value of the 1100 \micron\ observations is 0.12 M$_{\sun}$,
although this star was observed under poor atmospheric conditions. There
is a large NIR excess (Walter et al.  1994) suggesting the presence of
dust somewhere in this system. Jensen \& Mathieu (1997) argue that the
excess emission at 60 \micron\ suggests the presence of material outside
the binary orbit in a CB disk. Its SED is consistent with a CB disk
in which the central region has been cleared (Jensen \& Mathieu 1997).
With the mass function, inclinations lower than 22\arcdeg\ would give a
secondary mass higher than that of the primary, which is unlikely
(Jensen \& Mathieu 1997) so probably $i > 22\arcdeg$. S has a
prominent 10 \micron\ silicate emission feature, which is reproduced
with an optically thin inner disk. The continuous disk does not
reproduce the observations well. A gap or hole fit the data, but neither
does so perfectly (Jensen \& Mathieu 1997).

Data are presented in Table~\ref{Tab-n162819} and
Figure~\ref{Fig-n162819}. The average polarization, 2.92\% at 16\arcdeg,
is statistically variable. NTTS~162819-2423S is 0.5\arcdeg\ distant from
Haro~1-14C and 0.1\arcdeg\ from the previous star (see map in
Figure~\ref{Fig-sco5}). It has a lower polarization than the last star,
with 2.9\% average polarization at 16\arcdeg. Since it is
polarimetrically variable, it has an intrinsic polarization, but the IS
one dominates; there might also be an intra-cluster component. Since we
used a $8\farcs2$ aperture hole, it is not impossible that some
observations accidentally included more than just the southern
binary. We have made some tests using different aperture holes and
conclude that when the North and South components are included in the
measurements, the polarization is higher by $\approx 0.2$\%. That may
explain the chaotic polarimetric observations that show wild
fluctuations in polarization; nonetheless, some of the variations could
be attributed to the orbital motion of the southern component
alone. Future polarimetric observations should be made with much smaller
aperture holes, making sure what is included in the measurement.

\section{Orbital inclination \label{sect-incl}}
One of the goals of these polarimetric observations is to determine the
orbital inclinations of the selected PMS binaries, by using the BME
formalism. This formalism can still be used if the orbits are
non-circular and the scatterers are spherical grains, within the limits
presented in Papers~I and II.

Noise with a standard deviation greater than 10\% of the amplitude of
the polarimetric variations will prevent the BME formalism from finding
a reasonable estimate of the true inclination (Paper~I). Other studies
have also shown that the quality of the data (number of data points,
observational errors, amplitude of the polarimetric variations) can also
strongly influence the results found by the BME formalism.

Aspin, Simmons, \& Brown (1981) have studied the standard deviation
$\sigma_{\rm nec}(i)$ which is necessary to determine an inclination $i$
to $\approx \pm 5 \arcdeg$, with a 90\% confidence level. They give an
approximate relation for the data quality $DQ$:\\
\begin{equation}
DQ = \frac{\sigma_{\rm o}}{A_{\rm o} \sqrt{N_{\rm o}}} =
\frac{\sigma_{\rm nec}(i)}{A(i) \sqrt{N}}, 
\end{equation}
where 
\begin{equation}
A = \frac{|Q_{\rm max} - Q_{\rm min}| + |U_{\rm max} - U_{\rm min}|}{4},
\label{EQ-A}
\end{equation}
$\sigma_{\rm o}$ is the observational error of the polarization, $A_{\rm
o}$ is the observed polarimetric variability calculated with
Equation~\ref{EQ-A}, $N_{\rm o}$ is the number of observations, and
$N=40$ (the number of bins in their simulations). A set of very good
quality observations will have a low value of $DQ$. We present in
Table~\ref{Tab-noiseBME}, Column 2, $DQ$ values for the binaries studied
here. After the quantity $\sigma_{\rm nec}(i) / A(i)$ is calculated,
Table 1 in Aspin et al. (1981) gives the lowest possible inclination
that can be determined from the observations with a $\pm\ 5 \arcdeg$
accuracy at a significance of 10\% (meaning that the true inclination
has a probability of 90\% to be within $\pm$5\arcdeg\ of the value
returned by the BME formalism). If we apply this method to our sets of
data, we find that the quality of our data do not allow us to find $i$
to $\approx \pm 5 \arcdeg$, with a 90\% confidence level, for any of our
binaries.

Wolinski \& Dolan (1994) have also studied the confidence intervals for
orbital parameters determined polarimetrically. They made Monte Carlo
simulations of noisy polarimetric observations, for a specific geometry
not suitable for the stars studied in this present paper, but their
results are nonetheless instructive. Confidence intervals for $i$ are
given graphically as a function of a ``figure of merit'' $\gamma$:
\begin{equation}
\gamma = \left( \frac{A}{\sigma_p} \right) ^2 
\left( \frac{N}{2} \right),
\end{equation}
where $\sigma_p$ is the standard deviation of the noise that was added
to the data, $N$ is the number of observations, and $A$ is still given by
Eq.~\ref{EQ-A}. We have calculated
and present in Column 3 of Table~\ref{Tab-noiseBME} the figures of
merit $\gamma$ for some PMS binaries, by using the observational error
$\sigma(P)$ instead of the $\sigma_p$ used by Wolinski \& Dolan. It is
again seen that the quality of the data is not very good, mostly because
the amplitude $A$ is rather low (between 0.02 and 0.10\% in general).

Finally, following our own studies of the effects of noise on the BME
results (Paper~I), we have calculated the noise for the Stokes
parameters $Q$ and $U$, by using the variance of the fit and the
amplitude of the polarimetric variations; these amplitudes are computed
from the maximum and minimum values of the observations, and not those
of the fit. These calculations are presented in Columns 4 and 5 of
Table~\ref{Tab-noiseBME}, where levels of noise below 10\% are
non-existent. Once again, this analysis shows that the polarimetric
observations of PMS stars are not of ``very good quality'', not because
of instrumental or observational problems, but because non-periodic
stochastic polarization variations and low-amplitude periodic variations
make the data rather noisy, hiding whatever periodic polarimetric
variations exist. Future observations should be obtained at a site
offering many consecutive clear nights to cover in one run the whole
orbital period.

A more interesting case will be presented in a future paper (Manset \&
Bastien, in preparation). AK~Sco was observed within a few (12) consecutive
nights, and shows polarimetric variations of greater amplitude.

Assuming the BME formalism can be used to analyze the polarimetric
variations of binary PMS stars, we have added in
Table~\ref{Tab-noiseBME} the results of the BME analysis for the orbital
inclination. Most of the inclinations are near 90\arcdeg, which cannot
be a real result. This is compatible with the above discussion on the
effects of the stochastic noise on the inclination analysis.

\section{Orientation of the orbital plane and moments of the
distribution of the scatterers} In addition to the orbital inclination,
the BME formalism returns $\Omega$, the orientation of the orbital plane
with respect to the plane of the sky, and moments of the distribution of
the scatterers which are used to measure the asymmetry with respect to
the orbital plane ($\tau_0 G$), and the degree of concentration towards
the orbital plane ($\tau_0 H$). It is generally expected that the
distribution will be symmetrical to and concentrated in the orbital
plane, so $\tau_0 H > \tau_0 G$. In Table~\ref{Tab-omegagammas}, we
present the values of $\Omega$, $\tau_0 H$, $\tau_0 G$, and the ratio
$\tau_0 H / \tau_0 G$. If the circumbinary disks of these binaries can
be imaged (with interferometric or adaptive optics techniques), their
orientation should be similar to $\Omega$, although the orbits are not
necessarily coplanar with the disks or envelopes. The values for $\tau_0
G$ and $\tau_0 H$ are similar to those we have found in numerical
simulations (Paper~I), and to the observed values for other types of
binaries (Bastien 1988; Koch, Perry, \& Kilambi 1989). In particular,
$\tau_0 H > \tau_0 G$ as expected, with ratios approximately from 1.0 to
3.5.

These parameters are also calculated using the coefficients of the fits
and the same assumptions (the scatterers are electrons, the orbits are
circular). However, single-periodic variations that are not present when
the scatterers are electrons and the orbits are circular, do appear in
our simulations when there are dust grains instead of electrons, when
variable optical depth effects are considered, or when the orbits are
eccentric (Papers~I and II). Therefore, the values of the parameters
calculated might not reflect an asymmetric configuration ($\tau_0 G$) or
a concentration toward the orbital plane ($\tau_0 H$). For example, we
have found, using the code presented in Papers~I and II, that even
though $\tau_0 G$ should be null for a perfectly symmetric
configuration, it will not be so if we have dust grains or consider
variable optical depth effects.

\section{Search for Correlations}
To study correlations between variations and orbital parameters (period
and eccentricity), the maximum and minimum values (excluding the
atypical observations) were used to calculate $\Delta P$, $\Delta
\theta$, $\Delta Q$, $\Delta U$, and ($\Delta Q + \Delta U$ for all the
binaries of this study), and compare them to the orbital period and
eccentricity; see Table~\ref{Tab-AmpVar}, Figures~\ref{Fig-Var_vs_logP},
\ref{Fig-Var_vs_ecc}, and \ref{Fig-DqDu_vs_LogpEcc}. Note that in these
figures, the quantity $\Delta \theta$ for some stars is not meaningful
since their polarizations are very low, and hence, their position angles
not well defined. There are no clear correlations between the
polarimetric variations and the orbital characteristics, although the
most variable PMS binaries have intermediate periods (10--100 d) and
intermediate eccentricities (0.3--0.5). The converse is not true, and
there are examples of binaries with intermediate period or eccentricity
which do not present high variations. One should not forget, however,
that orbital inclination strongly affects the variability; a binary
which should display a high level of variability because of, say, its
orbital eccentricity, might not do so because its orbital inclination is
masking the variations.

Important variability for long orbital periods is not expected, since
the angular dimension of the CS disk as seen by the other star decreases
as distance increases, making the variations in scattering angles
smaller, and thus the polarimetric variations themselves smaller; this
has been verified using the numerical codes presented in Papers~I and
II. Moreover, the long-period binaries with periods over $\approx 100$ d
have probably not been adequately sampled to assess the polarimetric
variability over a full orbit. These two reasons could explain the drop
in variability for periods longer than about 100 d. The most variable
binaries all have high orbital eccentricities, $e>0.3$. The 2 stars with
high eccentricity but low-amplitude variations, Haro~1-14C and
NTTS~045251+3016, also have long periods (591 and 2530 d respectively),
which might explain why more variability is not seen.

Correlation between polarimetric properties and stellar age is difficult
to study because many factors introduce uncertainties in the derived
ages. First, the presence of unresolved companions, which increases the
total luminosity, can artificially lower the derived age (Simon et al.
1993). Second, the choice of evolutionary tracks affects the estimated
ages, as well as uncertainties in extinction or distance (Walter et
al. 1994; Forestini 1994; D'Antona \& Mazzitelli 1994). Third, Hartmann
et al. (1991) have suggested that prolonged disk accretion can arrest the
evolution so the true age is greater than the one found by the location
of a star in the HR diagram. So a binary can have a low polarization
because it is old and has lost its CS material, or because it is still
young but did not have much material to accrete due to an earlier SN
explosion for example. This discussion also does not take into account
the effects of inclination on the level of polarization. We could look
for correlations for binaries whose age are derived from the same
evolutionary tracks, but since many other factors influence the
polarization level, we have not tried to do so.

Our numerical simulations show that, as the orbital eccentricity
increases, variations go from being double-periodic (seen twice per
orbit) to being single-periodic (seen only once per orbit). Therefore,
as $e$ increases, the ratio of single-periodic over double-periodic
variations (1$\lambda$/2$\lambda$) should start from zero and
increase. However, the inclination also plays a role in the amplitude of
the variations and can decrease them. Table~\ref{Tab-lambda} and Figure~\ref{Fig-RatioLambdaE} show
this ratio for $P$, $\theta$, $Q$, and $U$, and all the binaries we have
studied. No correlation is seen, although this might be due to the
presence of noise in the observations. One of the highest-eccentricity
binaries, NTTS~162814-2427 ($e$=0.48) does present strong
single-periodic variations in $\theta$ and $U$, which is consistent with
the results of our numerical simulations. Other factors, though,
can produce single-periodic variations: geometry, variable absorption
effects.

\section{Discussion and summary}
We have presented polarimetric observations obtained at 7660\AA\ and
5550\AA\ for 7 spectroscopic PMS binaries located in the $\rho$~Oph and
Upper Sco star forming regions (SFRs). All binaries have detectable
linear polarizations and most are significantly affected by interstellar
(IS) polarization. NTTS~162814-2427 and NTTS~162819-2423S have high
intrinsic polarization and are 0\fdg2 from one another (0.22~pc at a
distance of 125~pc), so intra-cluster or very localized polarization
might also contribute to the observations. Since it does not affect
Haro~1-14C located 0\fdg5 away (see below), this localized polarization
would affect about 1~pc around NTTS~162814-2427 and NTTS~162819-2423S.
After an estimate of this IS polarization is removed, all binaries
present intrinsic polarizations $\gtrsim 0.4\%$ at 7660\AA. Two binaries
which, based on their SED and in particular their NIR excesses, have
detectable amounts of dust, NTTS~162814-2427 and NTTS~162819-2423S, also
have the highest intrinsic polarization ($\approx 1.5 - $2.1\%), which
might include some intra-cluster polarization.  The amplitudes of their
variations (0.5--1.0\%) also indicate strong intrinsic polarizations.

Of the 22 T~Tauri and Herbig Ae/Be binaries studied so far (Papers~III
and IV, this paper), 68\% (15/22) have intrinsic polarization above
0.5\% at 7660\AA. Many of those binaries do not present any evidence
other than polarimetric for the presence of dust, indicating that 
polarimetric techniques are more sensitive to the presence of dust
than photometric or spectroscopic ones. Most of the binaries with
intrinsic polarizations below 0.5\% are located in the Tau, Aur,
Ori SFRs, suggesting that these regions might be older than the
Sco and Oph ones.

All binaries located in the Sco and Oph SFRs are statistically variable
or possibly variable.When considering the binaries previously studied in
Papers~III and IV, and for which we have enough observations, including
data from the literature ($N$=21), 53\% (11/21) are clearly variable,
and 90\% (19/21) are variable or possibly so. These numbers include
variability seen in the periodic variations but not detected by the
statistical tests, and agree with results for single stars (Bastien
1988, M\'enard \& Bastien 1992). All the known CTTS binaries in our
sample (except DQ~Tau for which we only have one measurement) and many
NTTS and WTTS present polarimetric variations.

%

Three stars of the Oph and Sco SFRs have shown, twice, and at 7660\AA\
atypical values of polarization and/or position angle that are well
below or above the rest of the data (NTTS~155913-2233, NTTS~162814-2427,
NTTS~160905-1859). These atypical observations were also detected in our
previous samples (Par~1540, Par~2494, and MWC~1080). We believe these
are real observations of events that strongly affected the stars and/or
their environment.  For the stars with a sufficient amount of data,
additional observations at similar phases indicate that these atypical
points are not related to the normal periodic behavior. Interestingly,
in the Sco and Oph sample, only the shortest period binaries ($<$ 36~d)
have shown these atypical observations; the other 2 cases, with periods
of 89 and 144~d, did not show any such atypical observations. Although
atypical observations could possibly be detected with more observations,
this might indicate that the atypical observations are caused by the
close proximity of the stars, in particular, the interaction between
their CS disks.

A few of the PMS binaries present periodic polarimetric variations,
despite the noise introduced by non-periodic or pseudo-periodic
variations. Statistical tests conclude that NTTS~155913-2233's
polarization is possibly variable, and low-amplitude periodic variations
are seen when the data at 7660\AA\ are binned to minimize the noise.
NTTS~160814-1857 presents clear periodic and phased-locked variations
with an amplitude of 0.2\% over $\sim10$ orbits, and has not shown any
atypical observations, pointing to long-term stability. This might be
due to the period of 144.7 days, which keeps the stars and their CS
disks apart despite the relatively high eccentricity (0.26).
NTTS~160905-1859 presents periodic but not phased-locked variations. No
periodic variation is seen when viewing all the data, but more regular
variations are seen when taking shorter time intervals that include only
a few orbits at a time. This points to rapid re-arrangement of the CS
material, on a time scale of 1 orbit, and might also explain 2 atypical
observations. NTTS~162814-2427 presents very variable and noisy
polarization with an amplitude of 0.6\%. Whereas the variations as
usually seen in polarization, NTTS~162814-2427's variations are clear in
position angle, although with a low amplitude of less than
3\arcdeg. Moreover, the variations in position angle are mostly
single-periodic (seen once per orbit), which might be the result of the
high eccentricity 0.48, as predicted by our numerical simulations
(Papers~I and II). NTTS~162819-2423S presents high amplitude (1\%)
polarimetric variations, but those are very noisy. This might be due to
the quadruple nature of this system, and the accidental inclusion of
more than the South binary in some of the measurements. The amplitude of
the periodic polarimetric variations is greater in the Sco and Oph
sample ($0.2-1.2$\%) than in the Tau, Aur, Ori one ($\lesssim
0.3$\%). Additional polarimetric observations obtained over a shorter
period of time would be interesting to see if the noise is reduced and
the variations hence made clearer, and if the variations are still
periodic.

One of the goals of these observations was to find the orbital
inclinations. Unfortunately, non-periodic or pseudo-periodic variations
sometimes mask the truly periodic variations by introducing noise. This
noise is too high for the BME formalism to find reasonable estimates of
the orbital inclination, in the 22 binaries studied so far. Three
factors contribute to this difficulty. First, dust grains are the main
scatterers in these systems, and it has been shown that dust grains
produce polarimetric variations of smaller amplitude than electrons
(Paper~II). Second, the disk around these short-period binaries are
probably CB rather than CS ones, and CB disks produce variations of
smaller amplitudes than CS disks (Paper~II). Finally, non-periodic
events introduce noise that mask the already small amplitude
variations. This last problem might be improved by taking data on
shorter periods of time.

The BME formalism also returns moments of the distribution of the
scatterers, used to measure the asymmetry with respect to the orbital
plane and the degree of concentration towards the orbital
plane. Although the assumptions used in the BME formalism (scattering on
electrons, circular orbits) are not met in the PMS binaries studied
here, the values returned for PMS binaries are of the same order of
magnitude as values for other types of stars and as for our simulations
($\sim$1--10$\times 10^{-4}$).

There are no clear correlations between the polarimetric variations and
the orbital characteristics. The most variable binaries all have high
orbital eccentricities, $e>0.3$. The 2 stars with high eccentricity but
low amplitude variations, Haro~1-14C and NTTS~045251+3016, also have
long periods, which might explain why more variability is not seen.

\acknowledgments

N. M. thanks the directors of the Mont M\'egantic Observatory for
granting generous time over many years. The technical support from the
technicians of the observatory, B. Malenfant, G. Turcotte, and F. Urbain
is duly acknowledged.  N. M. thanks the Conseil de Recherche en Sciences
Naturelles et G\'enie of Canada, the Fonds pour la Formation de
Chercheurs et l'Aide \`a la Recherche of the province of Qu\'ebec, the
Facult\'e des Etudes Sup\'erieures and the D\'epartement de physique of
Universit\'e de Montr\'eal for scholarships, and P. B. for financial
support. We thank the Conseil de Recherche en Sciences Naturelles et
G\'enie of Canada for supporting this research.  We thank Fran\c{c}ois
M\'enard for providing the Pic-du-Midi data. N. M. is Guest User,
Canadian Astronomy Data Centre, which is operated by the National
Research Council, Herzberg Institute of Astrophysics, Dominion
Astrophysical Observatory. 


\newpage

\scalebox{0.75}{\includegraphics{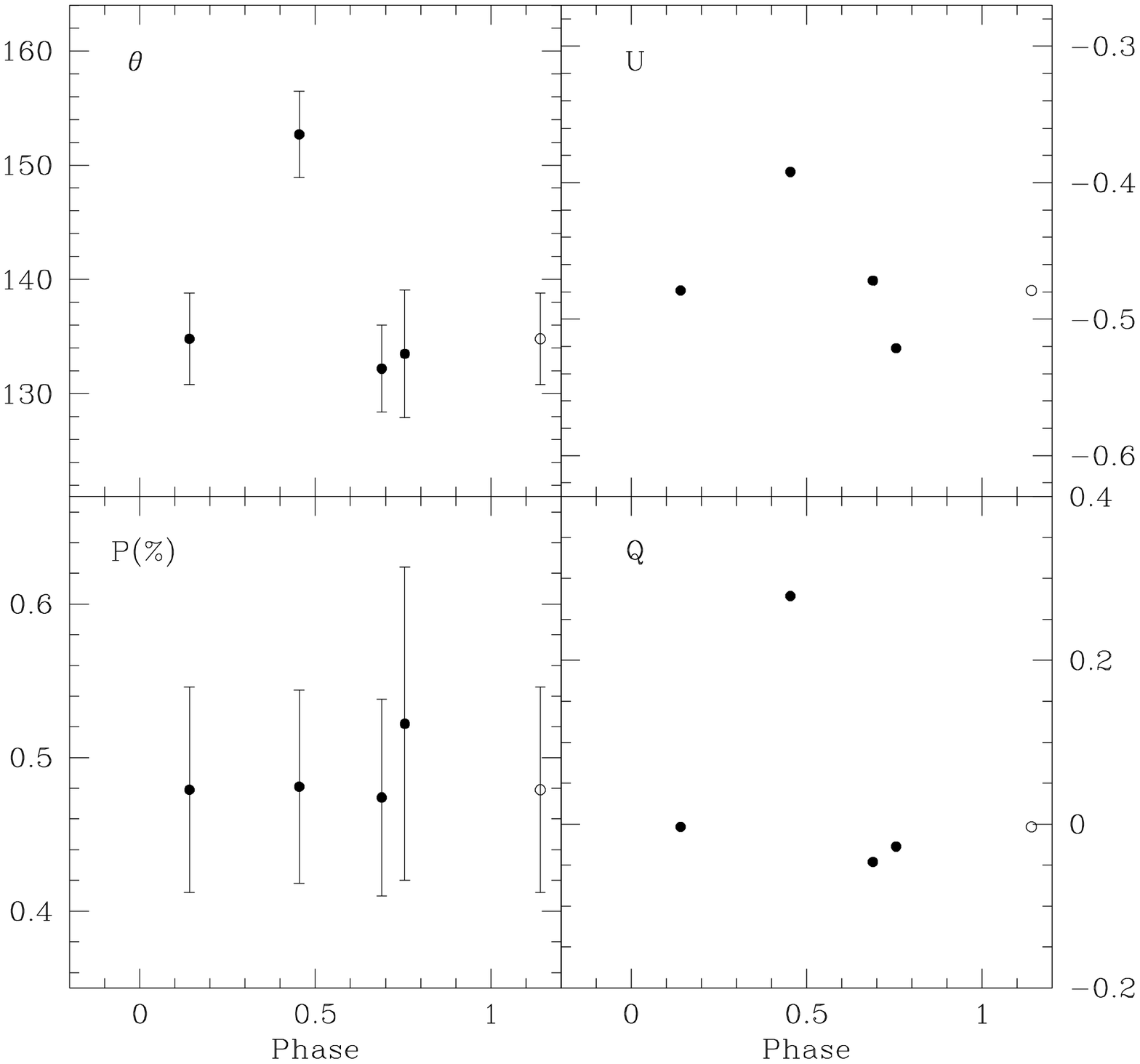}}
\figcaption[Manset5.fig01.ps]{Polarimetric observations of
NTTS~155808-2219. The polarization $P$, its position angle $\theta$, and
the Stokes parameters $Q=P \cos 2\theta$ and $U=P \sin 2 \theta$ are
graphed as functions of the orbital phase. \label{Fig-n155808}}
 
\newpage
\scalebox{0.75}{\includegraphics{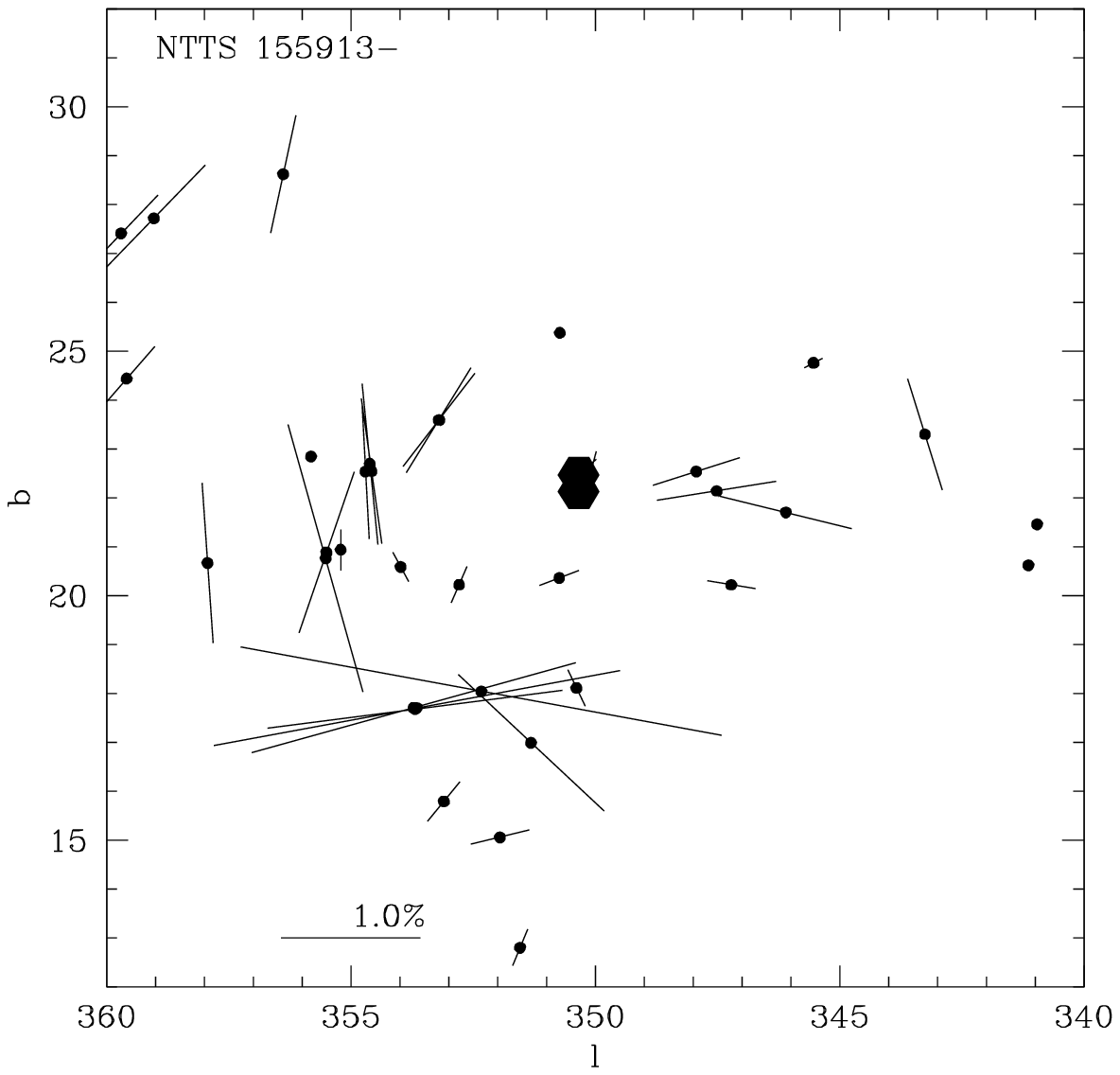}}
\figcaption[Manset5.fig02.ps]{Map of the interstellar polarization in the
vicinity of NTTS~155913-2233 (hexagonal symbol at the center of the map)
and NTTS~155808-2219 (above center). The stars selected to calculate the
IS polarization are within 80~pc of those targets. \label{Fig-sco3}}

\newpage
\scalebox{0.75}{\includegraphics{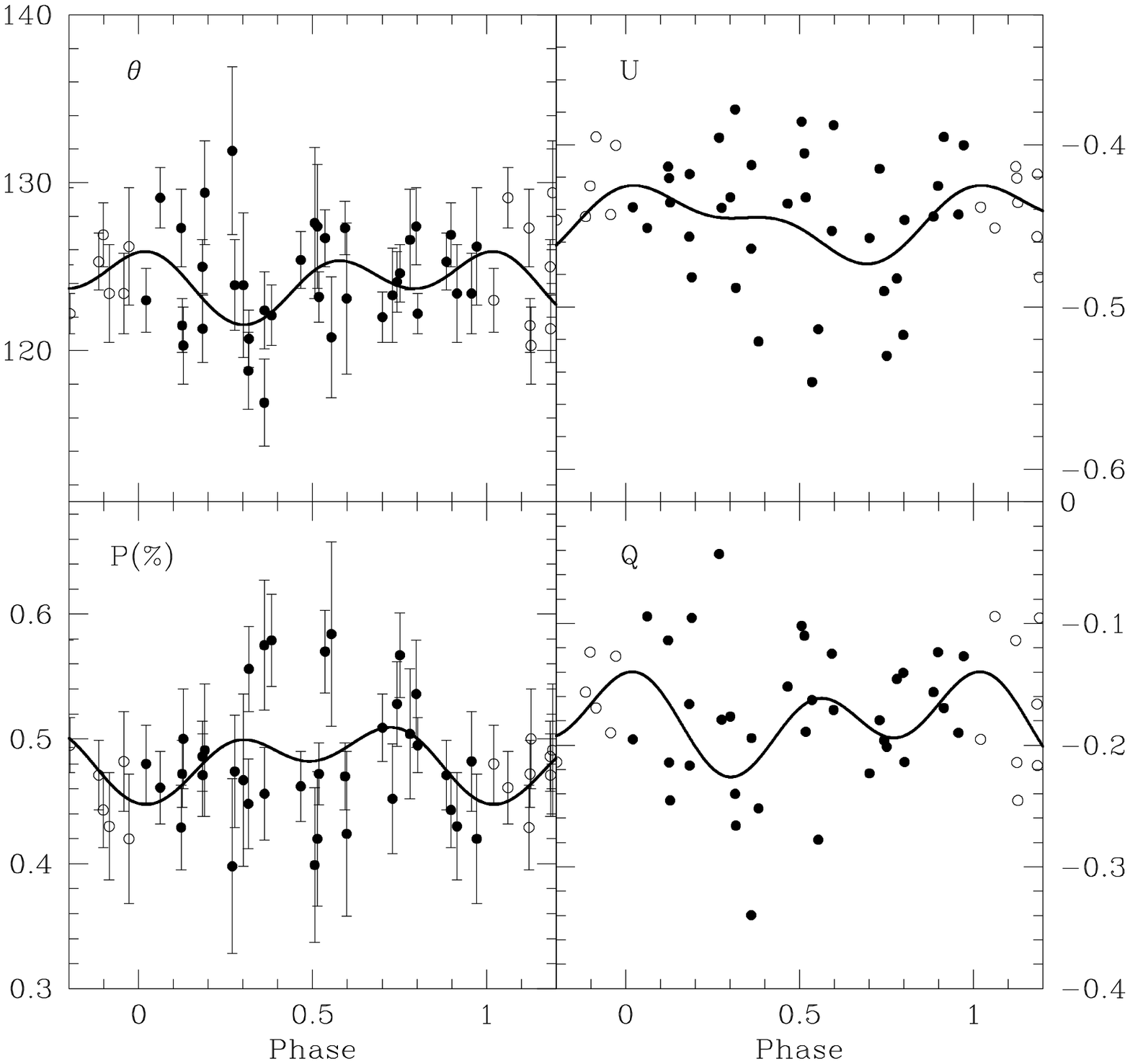}}
\figcaption[Manset5.fig03.ps]{Polarimetric observations of
NTTS~155913-2233. The star is variable polarimetrically, but there is a
lot of scatter and no clear periodic variations. Two atypical
observations, taken in 1995 May and 1997 April are not shown since their
position angle, 162\arcdeg\ and 168\arcdeg, are very different from the
rest of the observations.\label{Fig-n155913a}}

\newpage
\scalebox{0.75}{\includegraphics{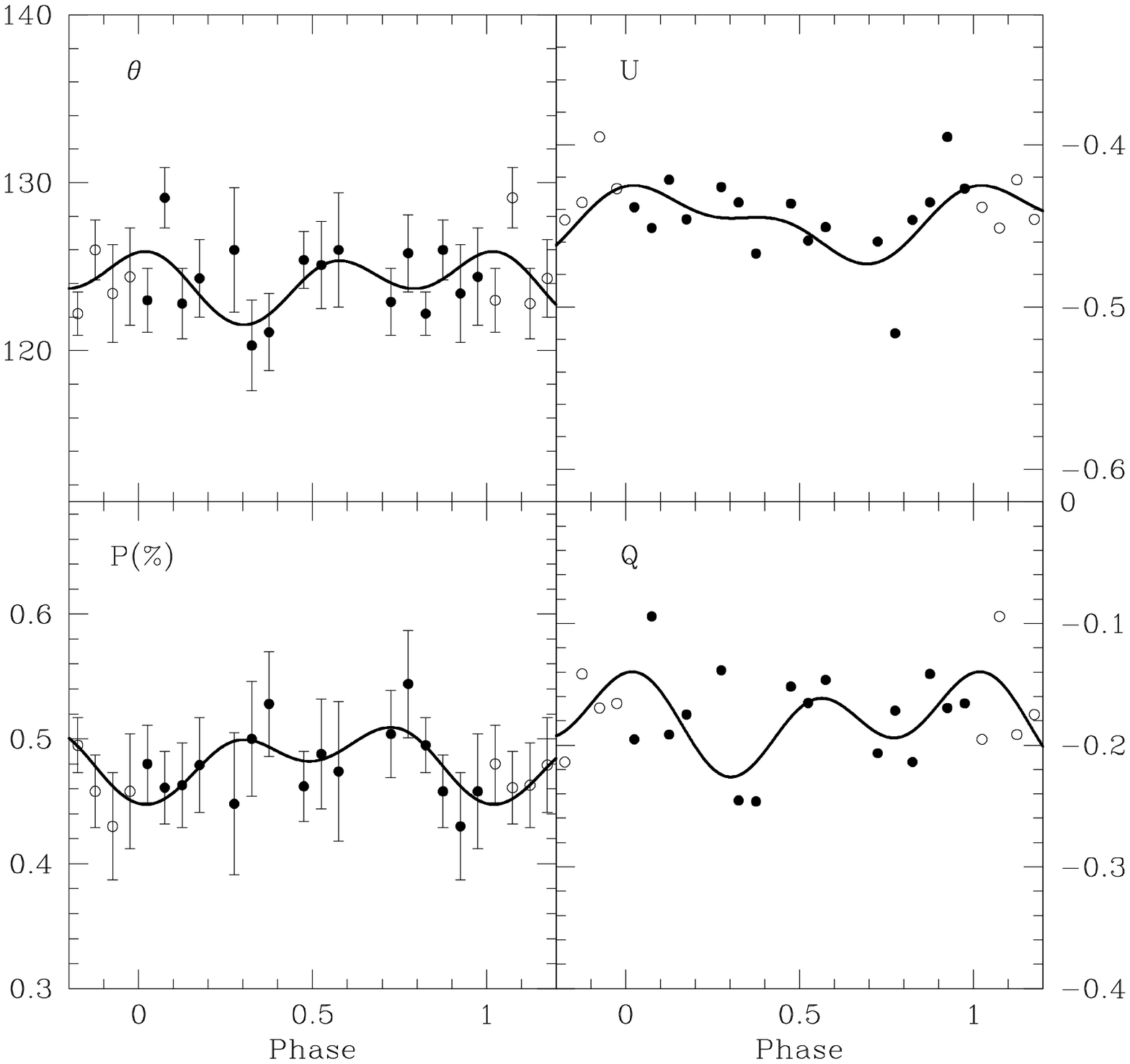}}
\figcaption[Manset5.fig04.ps]{Binned polarimetric observations of
NTTS~155913-2233; if there are periodic polarimetric variations, they
are of very small amplitude.\label{Fig-n155913b}}

\newpage
\scalebox{0.75}{\includegraphics{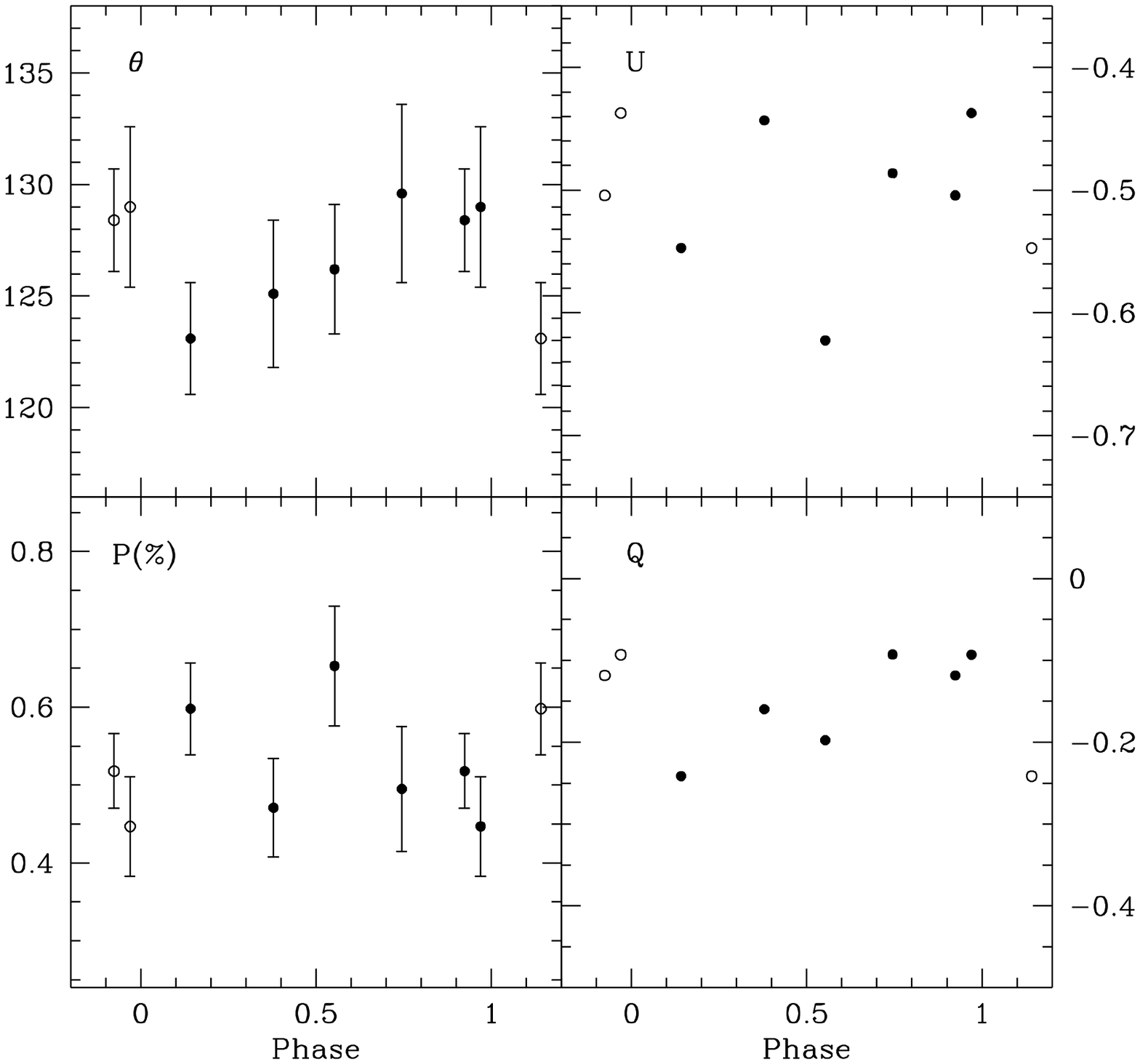}}
\figcaption[Manset5.fig05.ps]{Polarimetric observations for NTTS~155913-2233
obtained at 5550\AA. The position angle exhibits a systematic
variation. \label{Fig-pic155913}}

\newpage
\scalebox{0.75}{\includegraphics{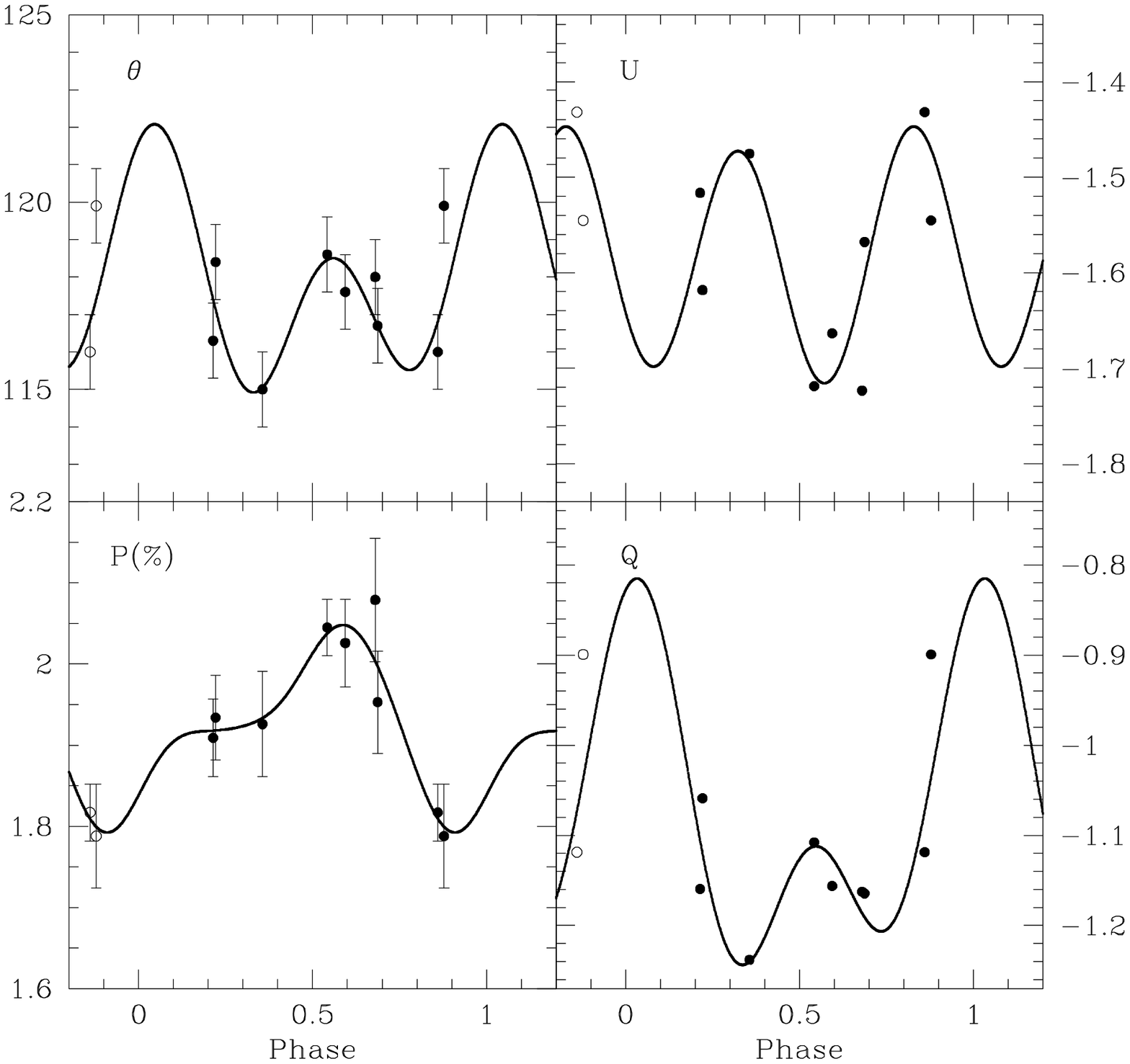}}
\figcaption[Manset5.fig06.ps]{Polarimetric observations of
NTTS~160814-1857.\label{Fig-n160814}}

\newpage
\scalebox{0.75}{\includegraphics{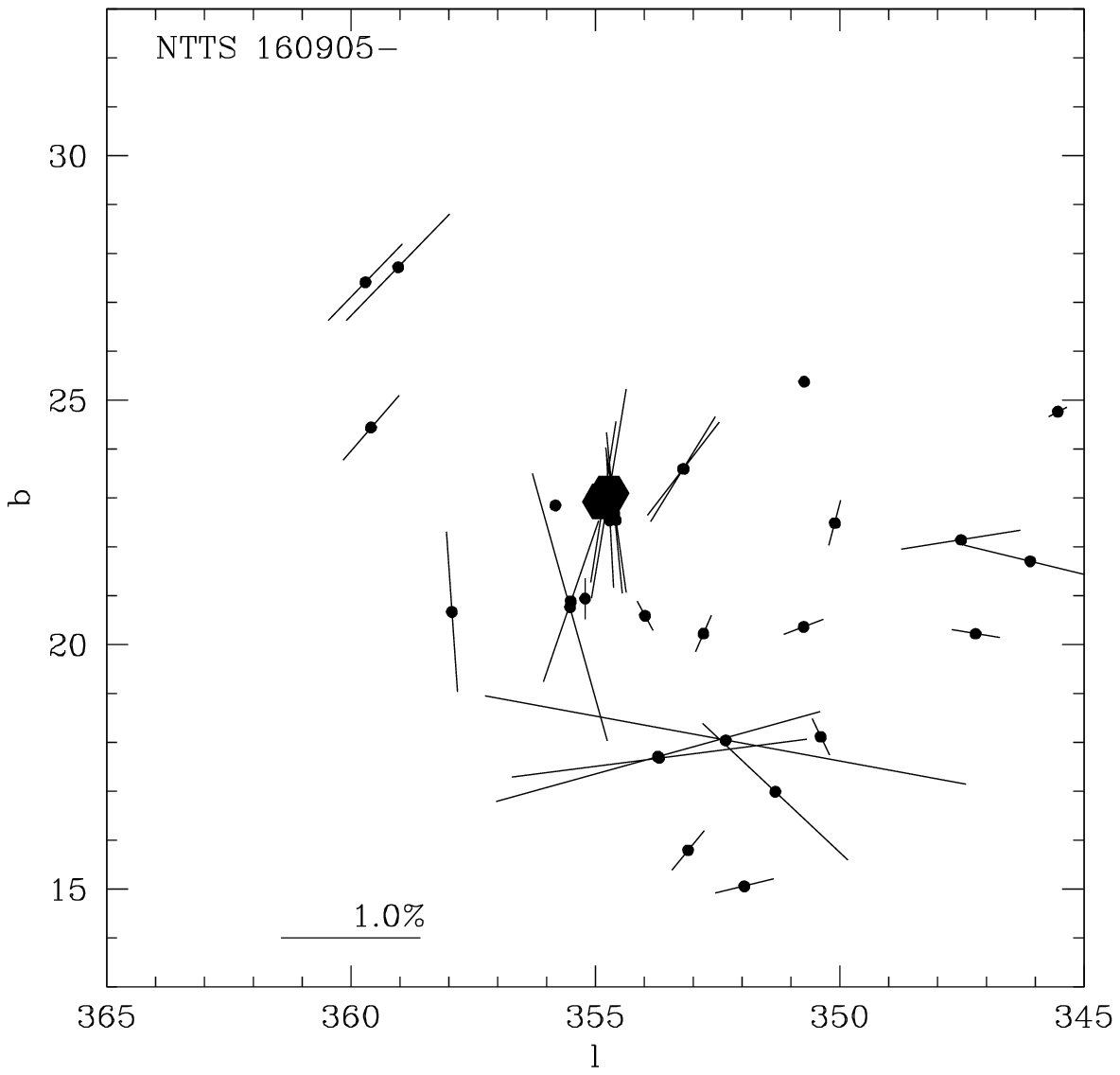}}
\figcaption[Manset5.fig07.ps]{Map of the interstellar polarization in the
vicinity of NTTS~160905-1859 and NTTS~160814-1857 (both at the center of
the map). The stars selected to calculate the IS polarization are within
75~pc of those targets. \label{Fig-sco4}}

\newpage
\scalebox{0.75}{\includegraphics{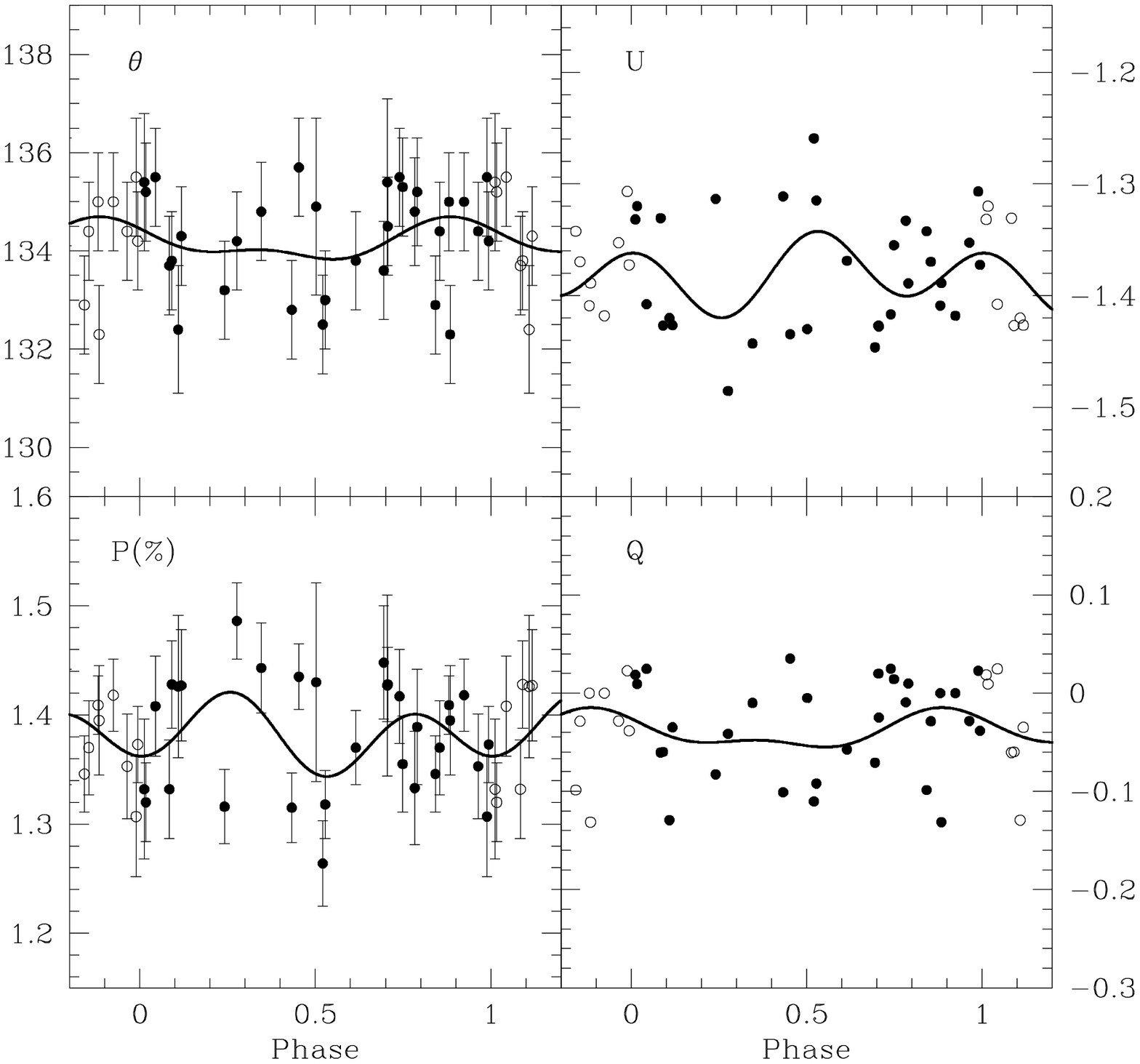}}
\figcaption[Manset5.fig08.ps]{Polarimetric observations of
NTTS~160905-1859 that show a lot of scatter. Two atypical observations,
taken in 1996 May and 1998 May, were removed since their position angle
was much lower or higher than the rest of the
observations.\label{Fig-n160905a}}

\newpage
\scalebox{0.75}{\includegraphics{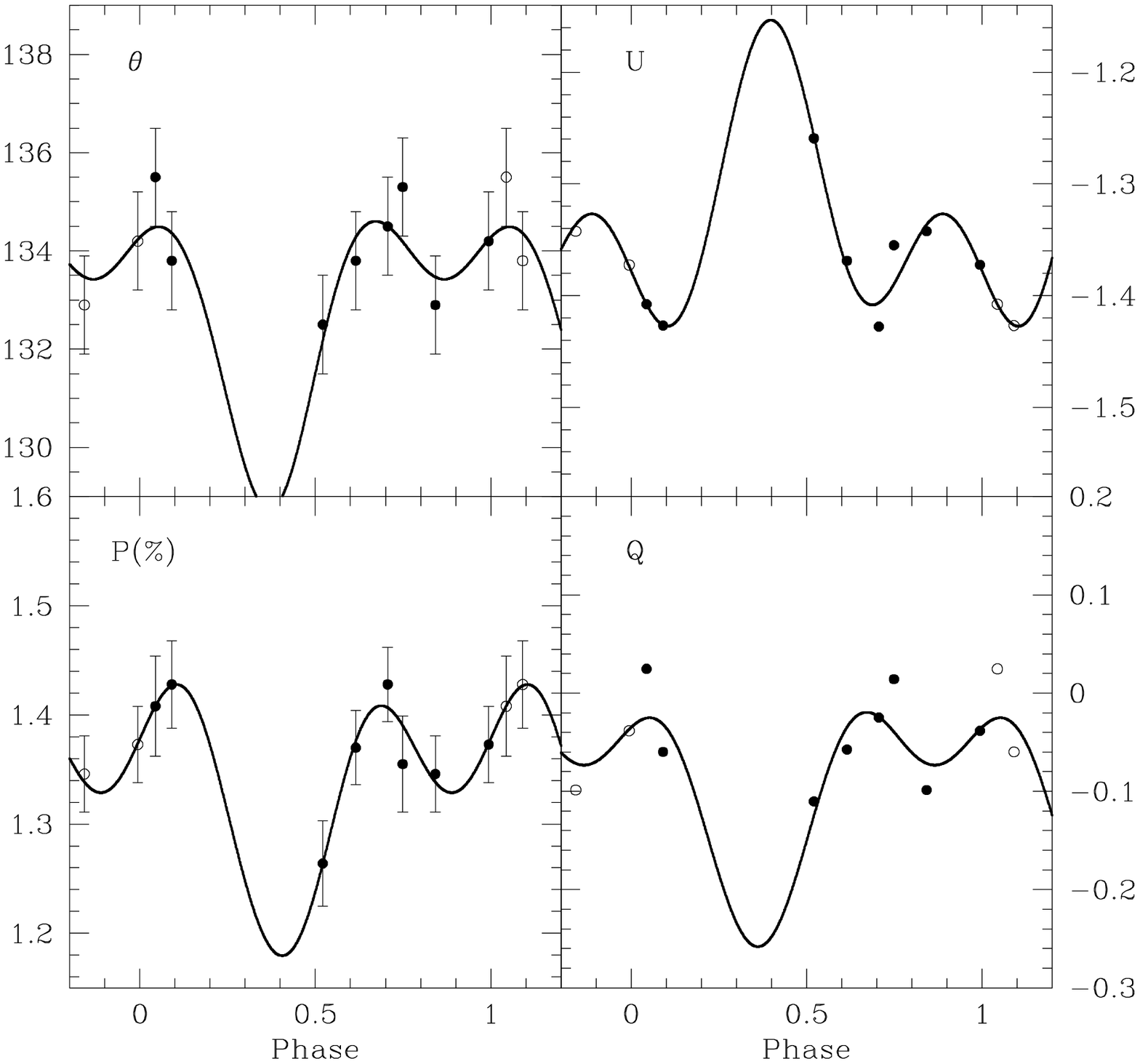}}
\figcaption[Manset5.fig09.ps]{Subset of the the polarimetric observations of
NTTS~160905-1859, obtained between 1997 April 2 and 1997 April
16.\label{Fig-n160905b}}

\newpage
\scalebox{0.75}{\includegraphics{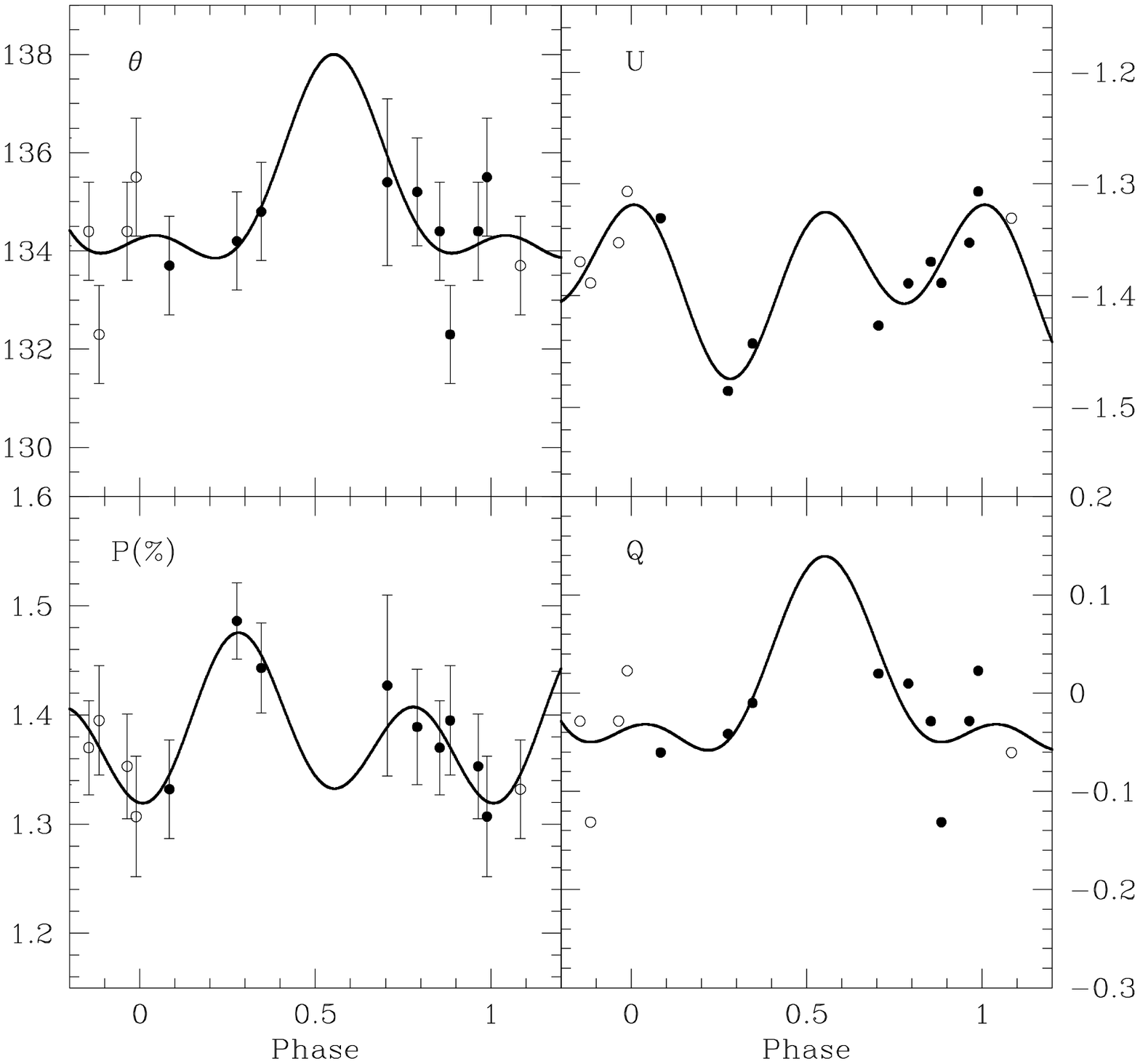}}
\figcaption[Manset5.fig10.ps]{Subset of the polarimetric observations of
NTTS~160905-1859 obtained between 1997 June 3 and July 11. The behavior
is clearly different from the one presented in the preceding
figure.\label{Fig-n160905c}}

\newpage
\scalebox{0.75}{\includegraphics{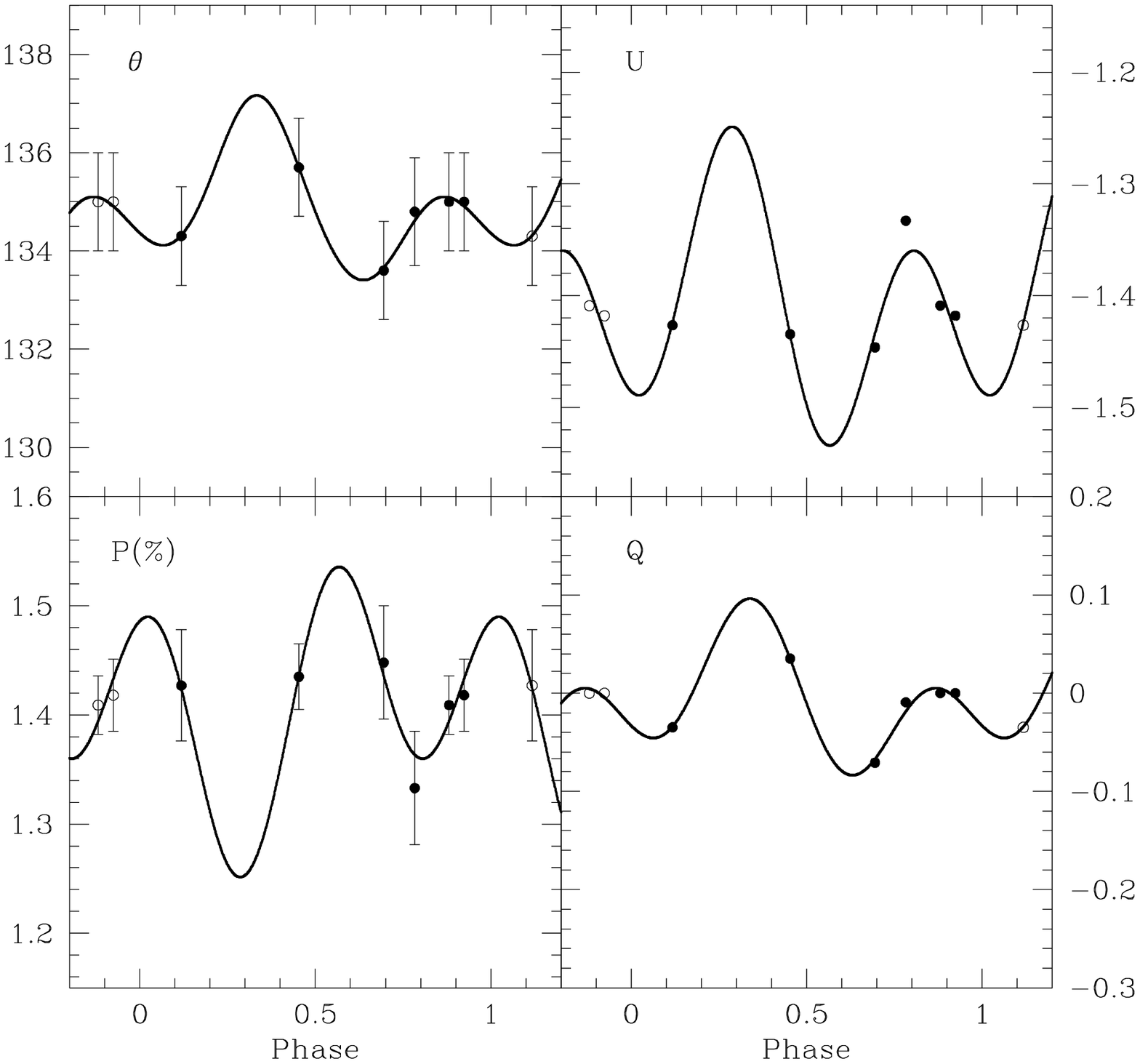}}
\figcaption[Manset5.fig11.ps]{Subset of the polarimetric observations of
NTTS~160905-1859, obtained between 1998 May 13 and June
1. \label{Fig-n160905d}}

\newpage
\scalebox{0.75}{\includegraphics{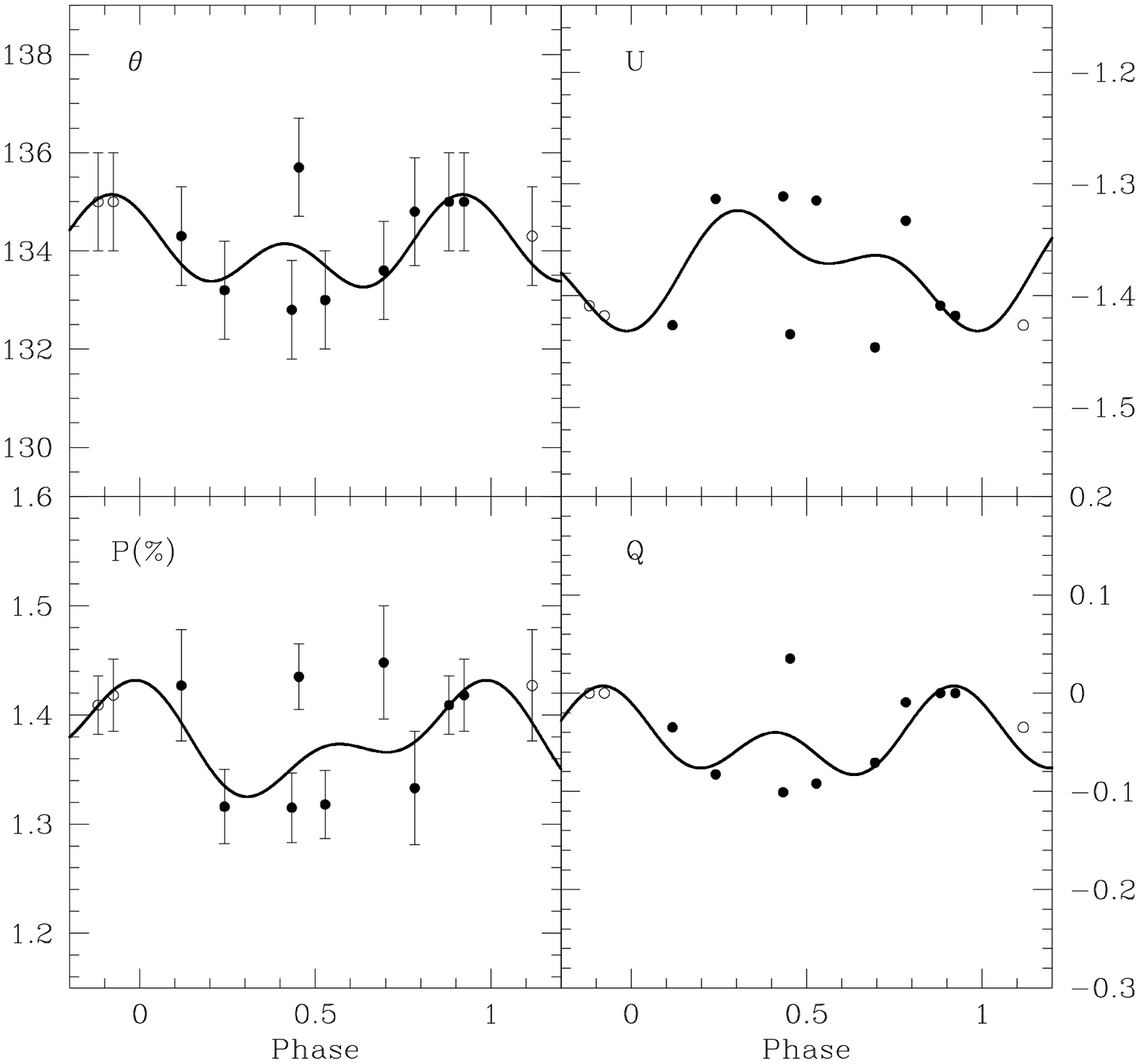}}
\figcaption[Manset5.fig12.ps]{Subset of the polarimetric observations of
NTTS~160905-1859, obtained between 1998 April 27 and June 1, thus
including the data presented in the preceding
figure.\label{Fig-n160905e}}

\newpage
\scalebox{0.75}{\includegraphics{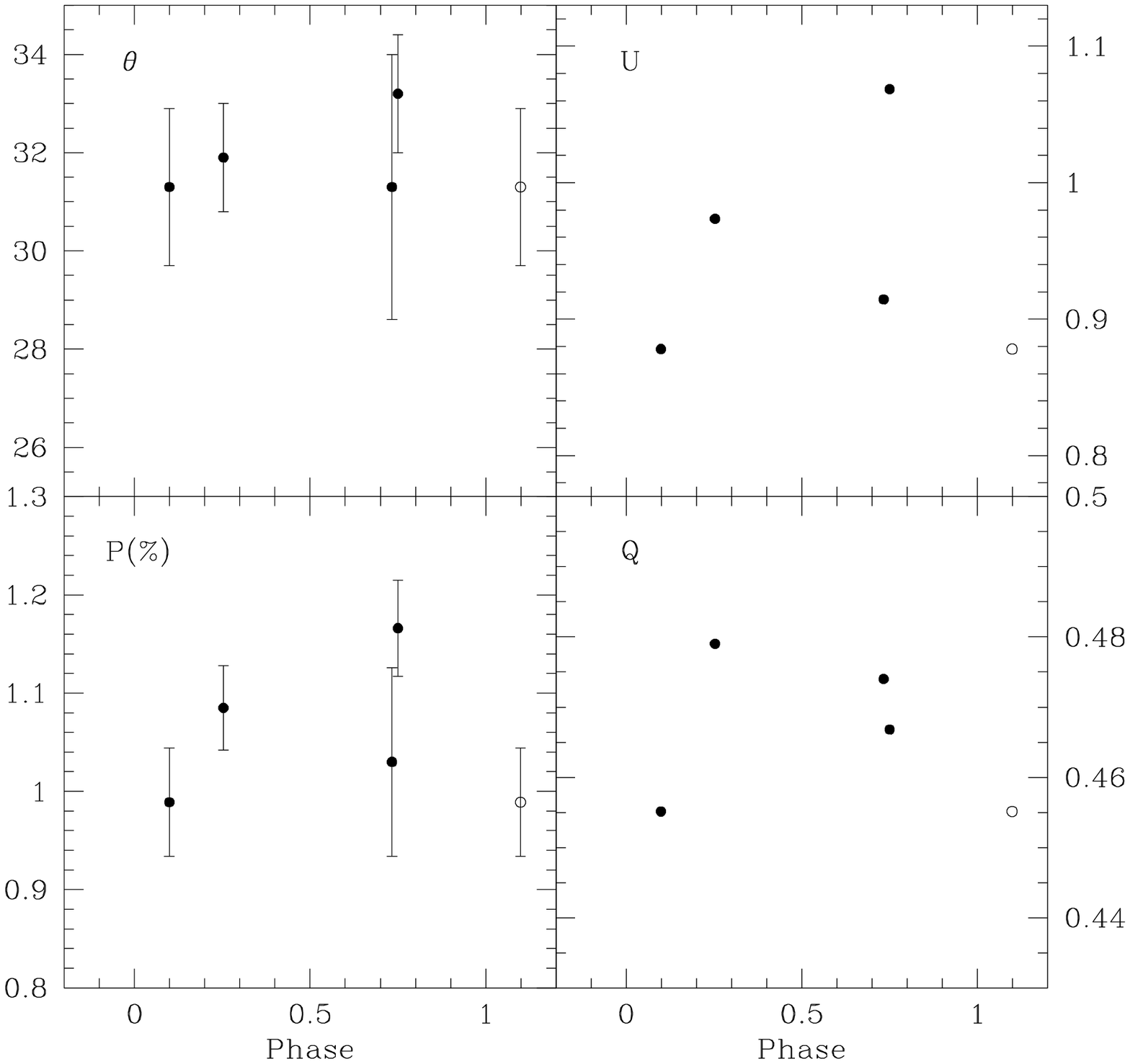}}
\figcaption[Manset5.fig13.ps]{Polarimetric observations of Haro~1-14C. One
atypical observation with a position angle different by 100\arcdeg\ is
not shown.\label{Fig-haro}}

\newpage
\scalebox{0.75}{\includegraphics{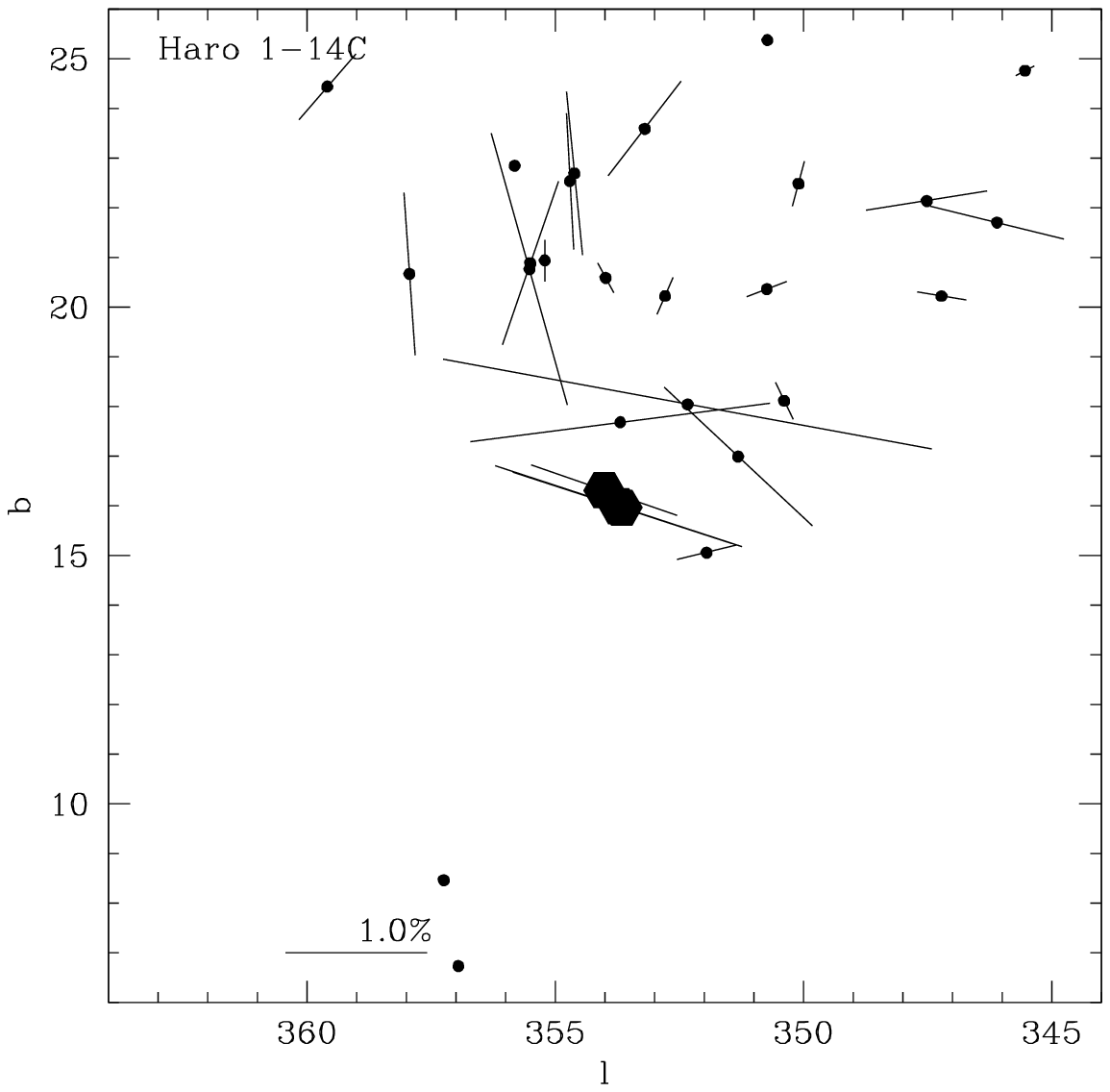}}
\figcaption[Manset5.fig14.ps]{Map of the interstellar polarization in the
vicinity of Haro~1-14C (at the center of the map), NTTS~162814-2427 and
NTTS~162819-2423S (both below center, at $\approx0\fdg5$ of
Haro~1-14C). The stars selected to calculate the IS polarization are
within 62~pc of those targets. \label{Fig-sco5}}

\newpage
\scalebox{0.75}{\includegraphics{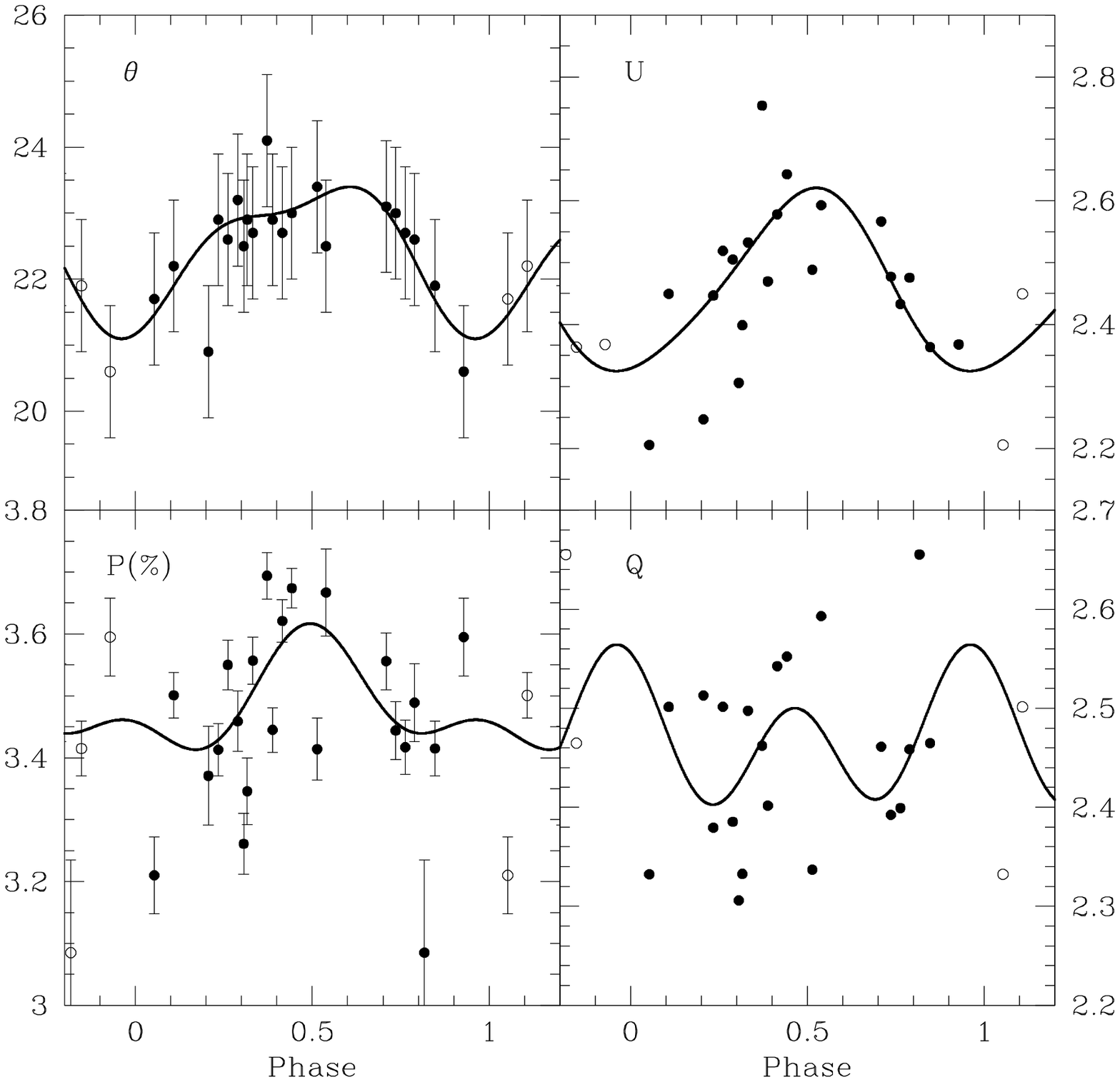}}
\figcaption[Manset5.fig15.ps]{Polarimetric observations of
NTTS~162814-2427. The star is clearly variable, but non-periodic
variations introduce a lot of scatter in the possible periodic
variations. One atypical point, taken in 1997 June, is not shown since
its position angle and polarization are below the rest of the
data.\label{Fig-n162814a}}

\newpage
\scalebox{0.75}{\includegraphics{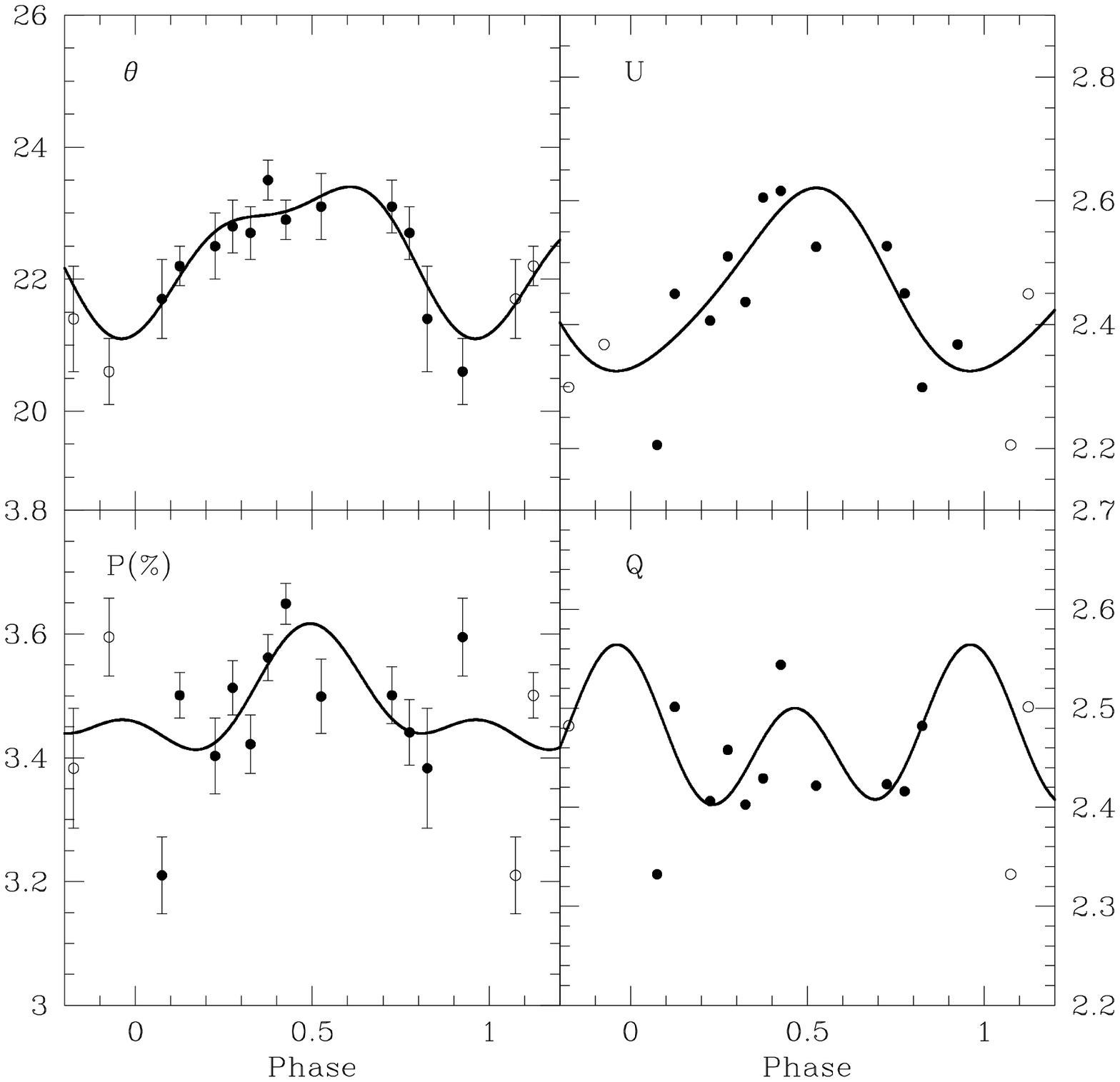}}
\figcaption[Manset5.fig16.ps]{Binned data for the polarimetric
observations of NTTS~162814-2427 that reveal clear single-periodic
variations in position angle and the $U$ parameter.\label{Fig-n162814b}}

\newpage
\scalebox{0.75}{\includegraphics{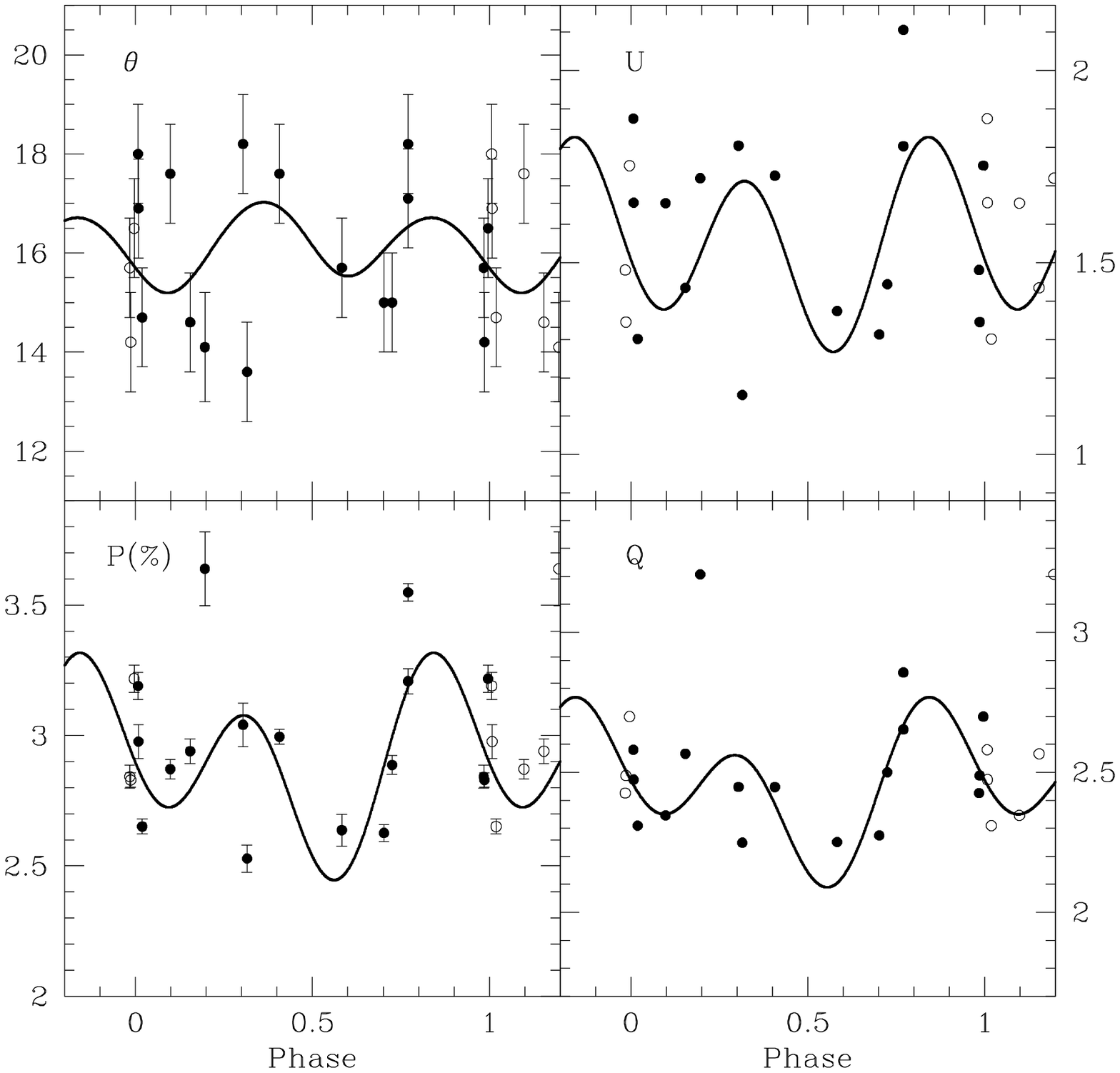}}
\figcaption[Manset5.fig17.ps]{Polarimetric observations of NTTS~162819-2423S,
the South binary of this quadruple system. This star is highly variable,
but some of the variability can be attributed to measurement that
included also the North component.\label{Fig-n162819}}

\newpage
\scalebox{0.75}{\includegraphics{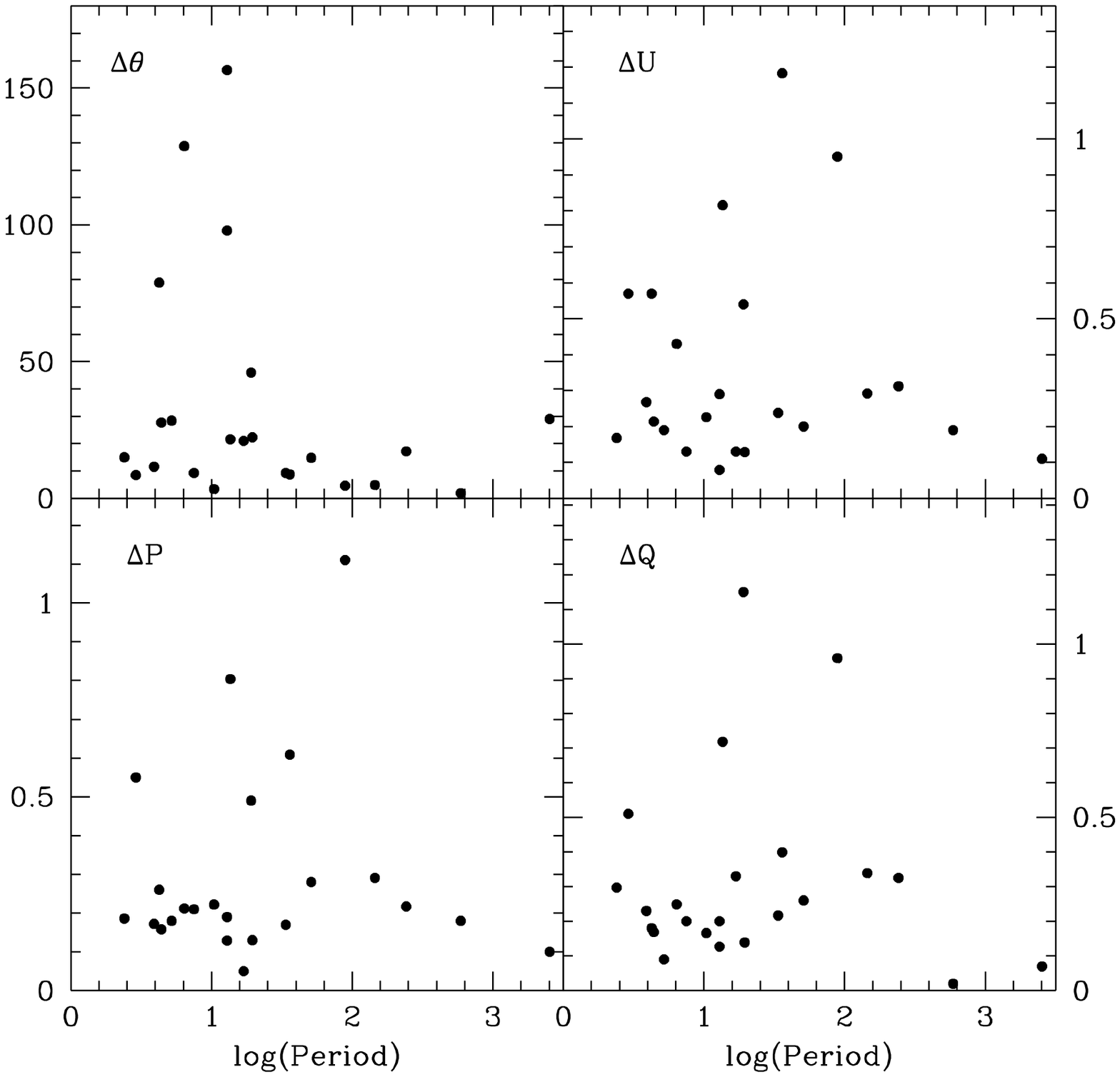}}
\figcaption[Manset5.fig18.ps]{Amplitude of the polarimetric variations as
a function of the orbital period. \label{Fig-Var_vs_logP}}

\newpage
\scalebox{0.75}{\includegraphics{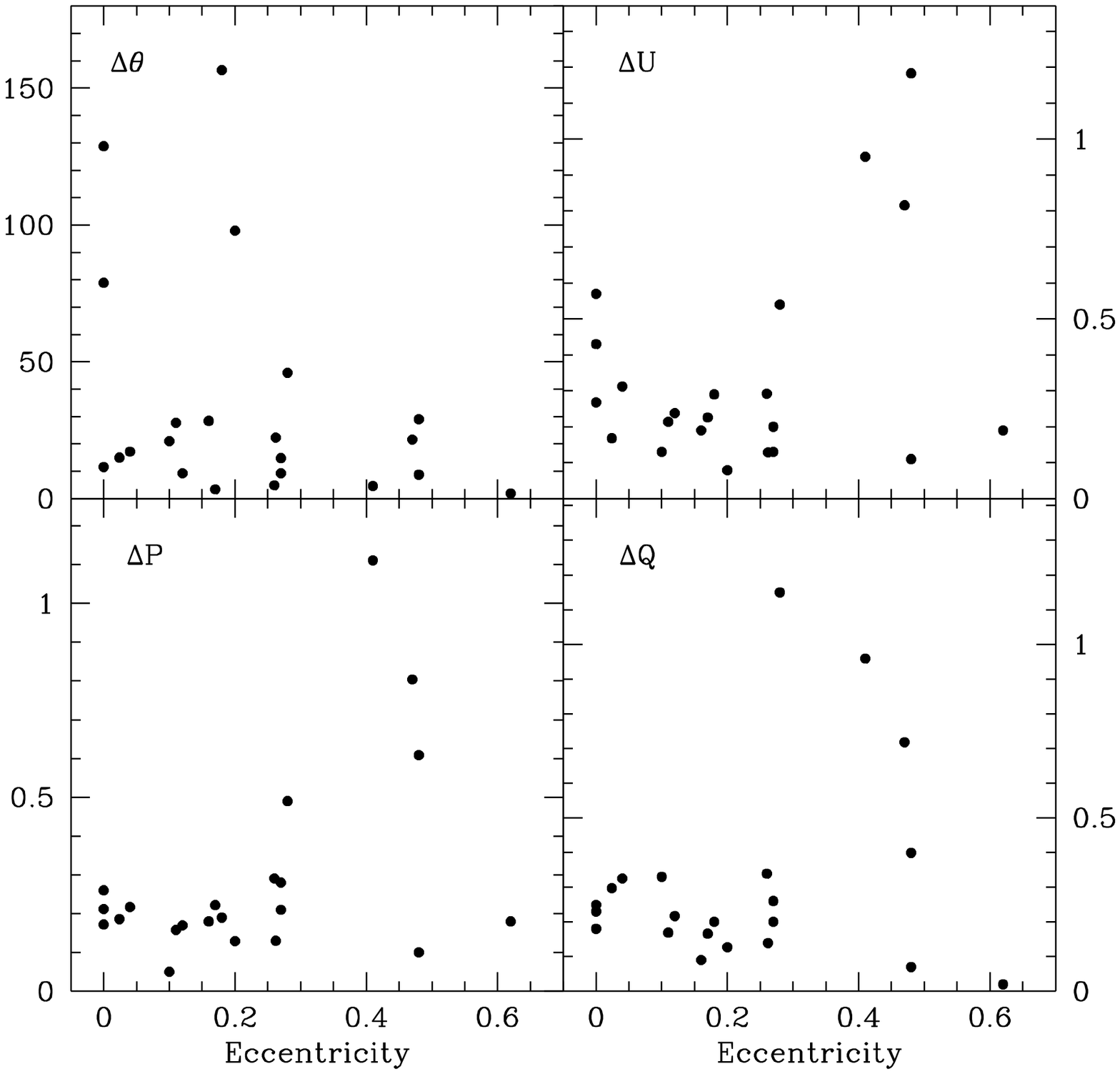}}
\figcaption[Manset5.fig19.ps]{Amplitude of the polarimetric variations as
a function of the orbital eccentricity. \label{Fig-Var_vs_ecc}}

\newpage
\scalebox{0.75}{\includegraphics{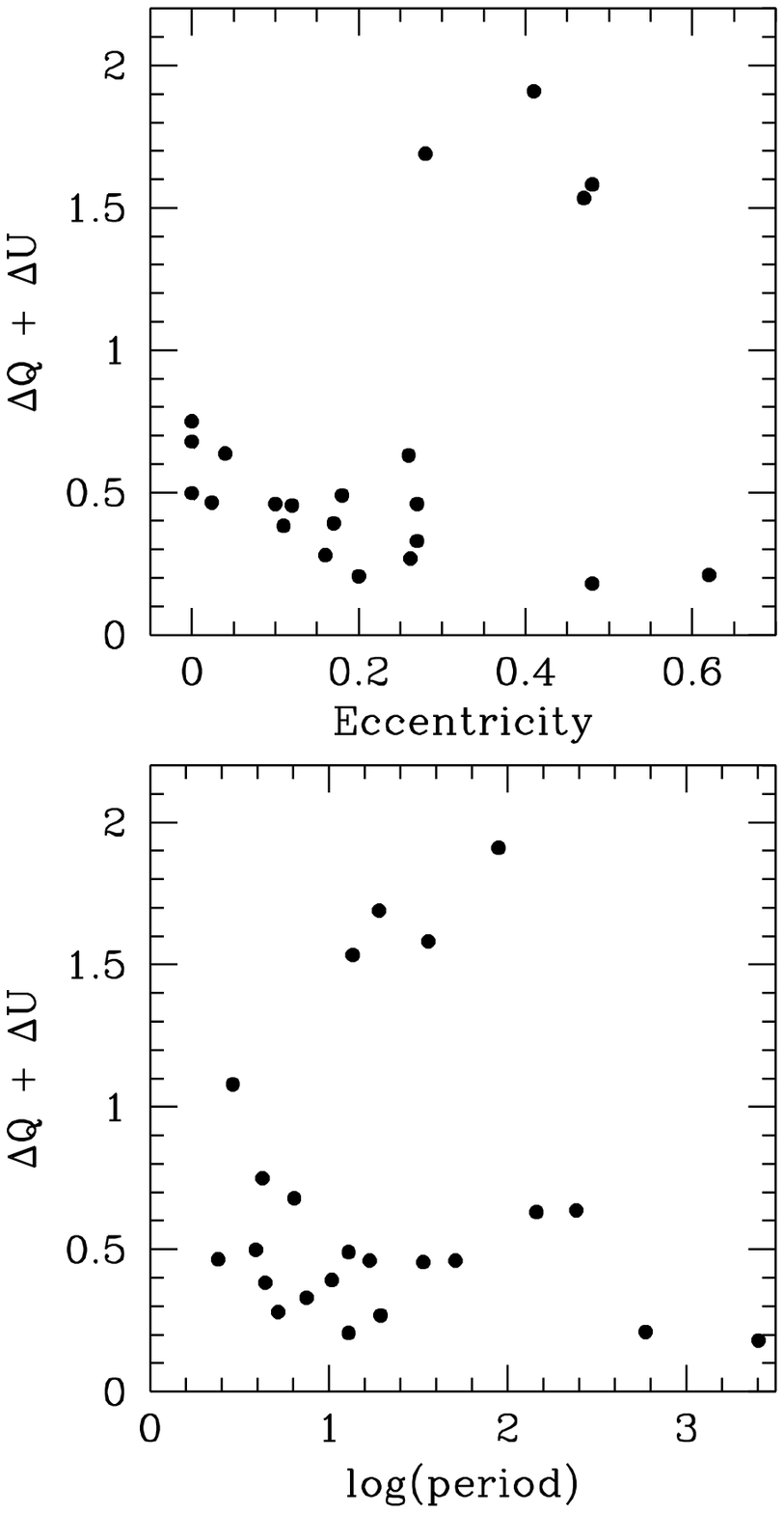}}
\figcaption[Manset5.fig20.ps]{Sum of the variations seen in the Stokes
parameters $Q$ and $U$ as a function of the orbital period and eccentricity.
\label{Fig-DqDu_vs_LogpEcc}}

\newpage
\scalebox{0.75}{\includegraphics{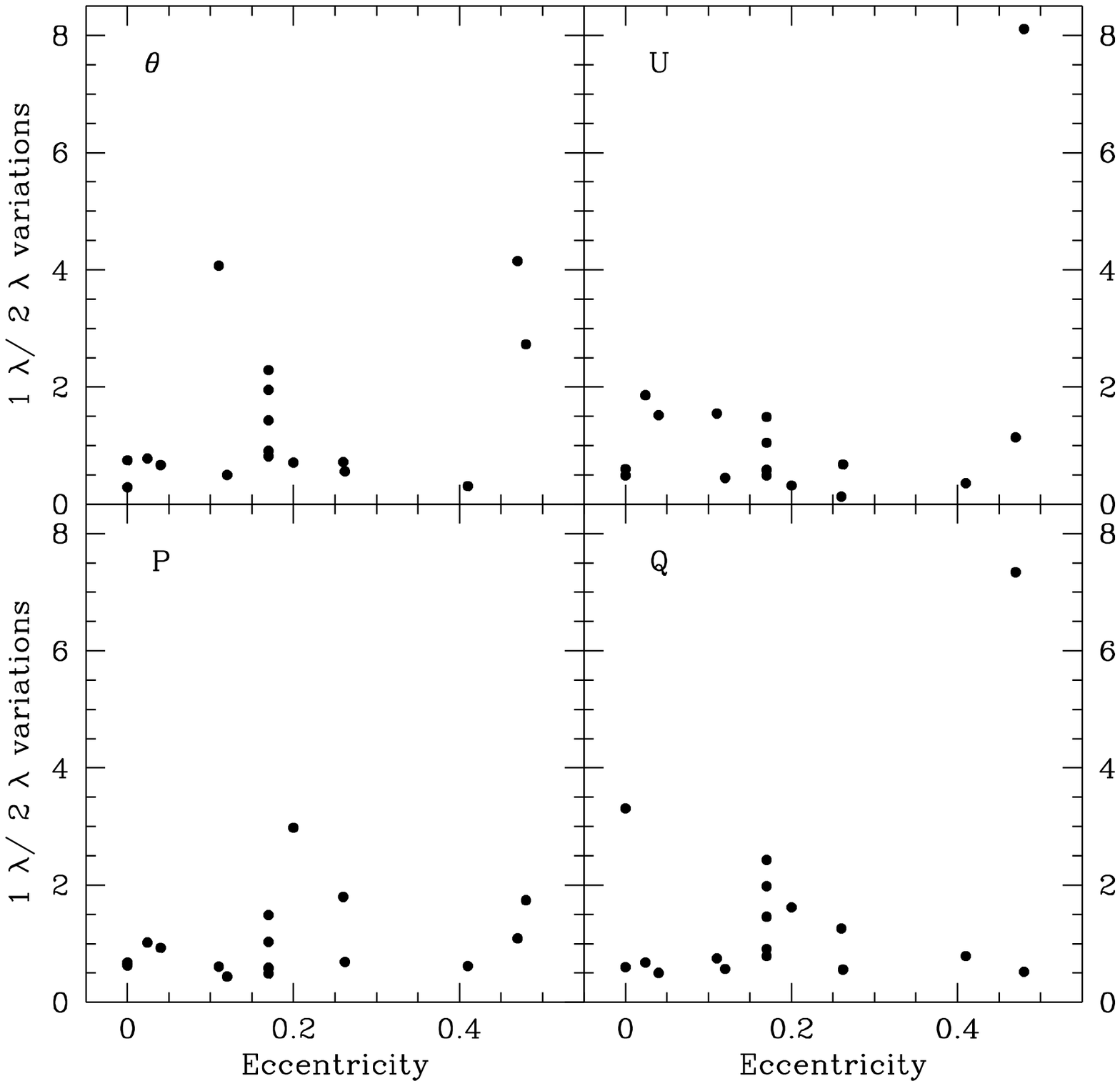}}
\figcaption[Manset5.fig21.ps]{Ratio of the amplitude of the
single-periodic variations over the amplitude of the double-periodic
variations. \label{Fig-RatioLambdaE}}


\newpage

\begin{deluxetable}{lrlccl}
\rotate         
\tabletypesize{\scriptsize}
\tablewidth{0pt}        
\tablecaption{Identification, coordinates, and location of the PMS
binaries \label{Tab-Coord}}
\tablehead{
\colhead{Star} & \colhead{HBC\tablenotemark{1}} & \colhead{Other Names} &
\colhead{$\alpha$\tablenotemark{2}} & \colhead{$\delta$\tablenotemark{2}} &
\colhead{Location}\\ 
\colhead{} & \colhead{} & \colhead{} & \colhead{(2000.0)} &
\colhead{(2000.0)} & \colhead{}}
\startdata
NTTS~155808-2219&\nodata & ScoPMS~20 & 16 01 06 & $-$22 27 00 & Upper Sco\\
NTTS~155913-2233&\nodata & ScoPMS~23 & 16 02 11 & $-$22 41 28 & Upper Sco\\
NTTS~160814-1857&630& ScoPMS~44, Wa~OPH~1, V1000~Sco & 16 11 09 & 
        $-$19 04 46 & Upper Sco (B40)\\
NTTS~160905-1859&633& ScoPMS~48, Wa~OPH~2, V1001~Sco & 16 11 59 & 
        $-$19 06 53 & Upper Sco (B40)\\
Haro~1-14C &644&\nodata & 16 31 05 & $-$24 04 40 & $\rho$ Oph B42\\
NTTS~162814-2427&\nodata & ROX~42 & 16 31 16 & $-$24 34 01 & $\rho$ Oph B42\\
NTTS~162819-2423S&\nodata & ROX~43A & 16 31 20 & $-$24 30 04 & $\rho$ Oph B42
\tablenotetext{1}{HBC numbers come from the Herbig and Bell Catalog
(Herbig \& Bell 1988).}
\tablenotetext{2}{All coordinates come from SIMBAD.}
\enddata
\end{deluxetable}

\pagebreak \clearpage


\begin{deluxetable}{llcccccrlcccrc}
\rotate
\tabletypesize{\scriptsize}
\tablewidth{0pt}
\tablecaption{Spectroscopic and orbital information for the PMS binaries
\label{Tab-Spectro}} 
\tablehead{
\colhead{Star\tablenotemark{1}} & \colhead{Spectral} & \colhead{Ref.} & 
\colhead{Type\tablenotemark{2}} & \colhead{Ref.} & 
\colhead{sgl/\tablenotemark{3}} 
& \colhead{Ref.} & \colhead{Period} & \colhead{Ecc.} & \colhead{Ref.} & 
\colhead{Inc.} & \colhead{Ref.} & \colhead{Dist.\tablenotemark{4}} &
\colhead{Ref.} \\ 
\colhead{}     & \colhead{Type} & \colhead{} & 
\colhead{} & \colhead{} &
\colhead{dbl} & \colhead{} &
\colhead{(d)} & \colhead{} & \colhead{} & 
\colhead{($\arcdeg$)} &\colhead{} & 
\colhead{(pc)} & \colhead{}}
\startdata
NTTS~155808-2219 (3) &\nodata &\nodata &\nodata &\nodata & dbl&1&16.925
& 0.10 &2&\nodata &\nodata&\nodata&\nodata\\ 
NTTS~155913-2233 (3) & K5 IV &3& WT&1& sgl&2&2.42378 & 0.024
&1&\nodata&\nodata&160&\nodata\\ 
NTTS~160814-1857 (2) & K2 IV + &3& WT &4& sgl&1&144.7 & 0.26
                     &2&\nodata &\nodata        &150&4\\ 
                     & M2-M3 & & &&&&&&&&&&\\
NTTS~160905-1859 (2) & K0 IV + &3& WT &4& sgl&1&10.400 & 0.17
                     &2&\nodata &\nodata&       150&4\\ 
                     & M3 IV & & & &&&&&&&&\\
Haro~1-14C (2) & K3 &5&\nodata &\nodata & sgl&2&591 & 0.62 &2&\nodata
                     &\nodata&125&\nodata\\ 
NTTS~162814-2427 (3)& K4-K5 &6& WT &6& dbl&1&35.95 & 0.48 &2
        & $\approx 71$ &7&125&1\\
NTTS~162819-2423S (2)& G8 &1&WT &1 & sgl&1&89.1 & 0.41 &2
        &$\sim60$ &7&125&\nodata
\tablenotetext{1}{The numbers in parenthesis after each object indicate
the number of stars known in each system.}
\tablenotetext{2}{Type of PMS star: WT (weak-line TTS)}
\tablenotetext{3}{Single-line (sgl) or double-line (dbl) spectroscopic
                     binary.}
\tablenotetext{4}{When no reference is indicated, we have used the known
                     distance to the cloud.}
\tablerefs{(1) Mathieu, Walter, \& Myers (1989);
(2) Mathieu (1994) and references cited;
(3) Walter et al. (1994);
(4) Walter (1986);
(5) Herbig \& Bell (1988) (HBC catalog) and references cited;
(6) Lee, Mart\'{\i}n, \& Mathieu (1994);
(7) Jensen \& Mathieu (1997)}
\enddata
\end{deluxetable}

\pagebreak \clearpage

\begin{deluxetable}{lccccccccccccc}
\rotate
\tabletypesize{\scriptsize}
\tablewidth{0pt}
\tablecaption{Average observed polarization, origin of the polarization,
and estimate of the intrinsic polarization for the PMS binaries
\label{Tab-averpol}}  
\tablehead{
\colhead{Star} & \colhead{$P_{\rm ave}$\tablenotemark{1}} & 
\colhead{$\theta_{\rm ave}$\tablenotemark{1}} & 
\colhead{$N_{\rm obs}$} & \colhead{Origin of} &
\colhead{$P_{\rm IS}$\tablenotemark{3}} & 
\colhead{$\sigma(P_{\rm IS})$} &
\colhead{$\theta_{\rm IS}$\tablenotemark{3}} & 
\colhead{$\sigma(\theta_{\rm IS})$ }&
\colhead{$N_{\rm IS}$} & \colhead{Radius} & \colhead{Distance} &
\colhead{$P_{\star}$\tablenotemark{4}} & 
\colhead{$\theta_{\star}$\tablenotemark{4}} \\

\colhead{} & \colhead{(\%)} & \colhead{(\arcdeg)} & 
\colhead{} & \colhead{Polariz-} &
\colhead{(\%)} & \colhead{(\%)} & 
\colhead{(\arcdeg)} & \colhead{(\arcdeg)} &
\colhead{} & \colhead{(\arcdeg)} & \colhead{Interval} &
\colhead{(\%)} & \colhead{(\arcdeg)}\\

\colhead{} & \colhead{} & \colhead{} &
\colhead{} & \colhead{ation\tablenotemark{2}} &
\colhead{} & \colhead{} & 
\colhead{} & \colhead{} & 
\colhead{} & \colhead{} & \colhead{($\pm$ pc)} & 
\colhead{} & \colhead{}}
\startdata
V773~Tau & 0.35 & 88 & 6  & IS $+$ $\star$ &
        0.92&0.20&72 &36&24&15&85&0.55&152\\
LkCa~3 & 0.05 & 76 & 12 & IS $+$ $\star$ &
        0.35&0.08&8  &18&23&15&70&0.35&95  \\
V826~Tau & 0.85 & 67 & 11 & $\star$ $+$ IS &
        0.21&0.04&131&32&32&15&80&0.97 &63 \\
UZ~Tau~E/W & 0.80 & 16 & 2  & IS $+$ $\star$  &
        0.68&0.16&17 &14&26&15&70&0.2 & 13\\
DQ~Tau & 0.57 & 79 & 1  & IS $+$ $\star$&
        0.35&0.09&85 &17&19&12&80&0.28& 72\\
NTTS~045251+3016& 0.10 & 107 & 3 & $\star$ $+$ IS &
        0.10&0.03&58 & 9&18&15&80&0.14&126 \\
GW~Ori& 0.61 & 126 & 11& $\star$ $+$ IS  &
        0.74&0.16&41 &30&30&10&200&1.26&129 \\
Par 1540 & 0.83 & 77 & 19 & IS $+$ $\star$  &
        1.37&0.25&64 &5&28&1.0&235&0.59&135\\
Par 2486& 0.14 & 63 & 6  & IS $+$ $\star$ &
        1.31&0.24&73 &5&32&1.0&235&1.02&164 \\
Ori~429 & 0.24 & 72 & 5  & IS $+$ $\star$  &
        1.01&0.23&92 &6&21&2&235&0.72& 8 \\
Par~2494 & 0.16 & 46 & 29 & IS $+$ $\star$  &
        1.10&0.17&78 & 5&48&2&235&0.91&173\\
Ori~569 & 0.18 & 76 & 4  & IS $+$ $\star$  &
        0.88&0.12& 90&5&68&6&235&0.62& 4 \\
W~134 & 0.22 & 32 & 11 & IS $+$ $\star$  &
        0.26&0.05&177&4&20&8&350&0.26& 60\\
VSB~126 & 0.16 & 66 & 6  & $\star$ $+$ IS  &
        0.26&0.05&177&4&20&8&350&0.36& 79\\
NTTS~155808-2219& 0.48 & 139 & 4 & $\star$ $+$ IS &0.34&0.07&88
&24&22&6&80&0.62&153\\
NTTS~155913-2233& 0.49 & 124 & 36& $\star$ $+$ IS &0.23&0.05&73
&26&22&6&80&0.57& 134\\
NTTS~155913-2233\tablenotemark{5}& 0.53 & 127 & 6&$\star$ $+$ IS&
0.23&0.05&73&26&22&6&80&0.64&137\\
NTTS~160814-1857& 1.94 & 117 & 9 & IS $+$ $\star$
&1.51&0.26&123&10&24&8&75&0.70& 105\\
NTTS~160905-1859& 1.38 & 134 & 34& IS $+$ $\star$
&1.16&0.20&123&10&24&8&75&0.58& 155\\
NTTS~160905-1859\tablenotemark{5}& 1.09 & 134 & 4 &IS $+$ $\star$&
1.16&0.20&123&10&24&8&75&0.43&178\\
Haro~1-14C      & 1.08 & 34 & 5  & IS $+$ $\star$  &1.08&0.21&24
&18&20&10&62&0.37& 64\\
NTTS~162814-2427& 3.50 & 22 & 22 & $\star$ $+$ IS  &1.62&0.32&25
&16&19&10&62&2.09& 20\\
NTTS~162819-2423S& 2.92 & 16 & 17& IS $+$ $\star$  &1.83&0.36&25
&16&19&10&62&1.47& 6
\tablenotetext{1}{Weighted averages of the observed $P$ and
$\theta$. The averages do not include atypical observations.}
\tablenotetext{2}{Probable origin of the polarization. A $\star$ symbol
indicates intrinsic polarization; IS stands for interstellar
polarization. If IS comes before a $\star$ symbol, the IS component of
the polarization is probably stronger than the intrinsic one.} 
\tablenotetext{3}{Estimate of the IS polarization at the location of the
PMS binary based on data from the Heiles (2000) catalog. See text for
details.} 
\tablenotetext{4}{Intrinsic polarization obtained after
subtracting the estimated IS polarization.}
\tablenotetext{5}{The additional entries for NTTS~155913-2233 and
NTTS~160905-1859 come from the Pic-du-Midi data taken in the $V$ filter.}
\enddata
\end{deluxetable}

\pagebreak \clearpage

\begin{deluxetable}{lcccccc}
\tablewidth{0pt}
\tablecaption{Amplitude of the polarimetric variations \tablenotemark{1}
\label{Tab-AmpVar}}
\tablehead{
\colhead{Star} & \colhead{Wavelength} & \colhead{$\Delta P$} &
\colhead{$\Delta \theta$} & \colhead{$\Delta Q$} & \colhead{$\Delta U$}
& \colhead{$N_{\rm obs}$}\\ 
\colhead{} & \colhead{(\AA)} &\colhead{(\%)} & \colhead{$(\arcdeg) $} &
\colhead{(\%)} & \colhead{(\%)} & \colhead{}} 
\startdata
V773~Tau	& 7660 & 0.28 & 15 & 0.26 & 0.20 & 6\\
LkCa~3		& 7660 & 0.13 & 98 & 0.13 & 0.08 & 12\\
V826~Tau	& 7660 & 0.17 & 12 & 0.23 & 0.27 & 11\\
UZ~Tau~E/W	& 7660 & 0.49 & 46 & 1.15 & 0.54 & 2\\
NTTS~045251+3016& 7660 & 0.10 & 29 & 0.07 & 0.11 & 3\\
GW~Ori		& 7660 & 0.22 & 17 & 0.33 & 0.31 & 11\\
Par~1540	& 7660 & 0.17 & 9  & 0.22 & 0.24 & 19\\
Par~2486	& 7660 & 0.18 & 28 & 0.09 & 0.19 & 6\\
Ori~429		& 7660 & 0.21 & 9  & 0.20 & 0.13 & 5\\
Par~2494	& 7660 & 0.13 & 22 & 0.14 & 0.13 & 29\\
Ori~569		& 7660 & 0.26 & 79 & 0.18 & 0.57 & 4\\
W~134		& 7660 & 0.21 & 129 & 0.25 & 0.43 & 11\\
VSB~126		& 7660 & 0.19 & 157 & 0.20 & 0.29 & 6\\
NTTS~155808-2219& 7660 & 0.05 & 21 & 0.33 & 0.13 & 4\\
NTTS~155913-2233& 7660 & 0.19 & 15 & 0.30 & 0.17 & 36\\
\nodata         & 5550 & 0.21 & 7 & 0.15 & 0.19 & 6\\
NTTS~160814-1857& 7660 & 0.29 & 5 & 0.34 & 0.29 & 9\\
NTTS~160905-1859& 7660 & 0.22 & 3 & 0.17 & 0.23 & 34\\
\nodata         & 5550 & 0.09 & 4 & 0.16 & 0.09 & 4\\
Haro~1-14C      & 7660 & 0.18 & 2 & 0.02 & 0.19 & 4\\
NTTS~162814-2427& 7660 & 0.61 & 9 & 0.40 & 1.18 & 22\\
NTTS~162819-2423S&7660 & 1.11 & 5 & 0.96 & 0.95 & 17\\
AK~Sco\tablenotemark{2}	& 5250 & 0.80 & 22 & 0.72 & 0.82 & 27\\
EK~Cep\tablenotemark{2}	& 5500 & 0.16 & 28 & 0.17 & 0.21 & 36\\
MWC~1080\tablenotemark{3}& 7660 & 0.55 & 9  & 0.51 & 0.57 & 62
\tablenotetext{1}{Difference between the minimum and maximum values of
the polarization, excluding, in some cases, atypical observations. The
values $\Delta Q$ and $\Delta U$ are not calculated from $\Delta P$ and
$\Delta \theta$, but from the maximum and minimum values of $Q$ and $U$.}
\tablenotetext{2}{Data for AK~Sco and EK~Cep will be presented in future
papers.}
\tablenotetext{3}{Data for MWC~1080 were presented in Paper~III. }
\enddata
\end{deluxetable}

\pagebreak \clearpage

\begin{deluxetable}{lcccccccc}
\tablewidth{0pt}
\tablecaption{Results of the variability tests at 7660\AA\  and 5550\AA
\label{Tab-VarDetails}} 
\tablehead{
\colhead{Star} & \colhead{Wavelength} & \colhead{$N_{\rm obs}$} &
\colhead{} &  \colhead{$\sigma_{\rm sample}$} & \colhead{$\sigma_{\rm
mean}$} & \colhead{$Z \pm \sigma_{Z}$} & \colhead{$P{\chi^2}$} &
\colhead{$P{\chi^2}$}\\
\colhead{} & \colhead{(\AA)} & \colhead{} & \colhead{} & 
\colhead{} & \colhead{} &
\colhead{} & \colhead{$1\sigma$} &
\colhead{$1.5\sigma$}}
\startdata
NTTS~155808-2219 &7660& 4&$Q$& 0.1531 & 0.0350 & 2.35 0.41 & 1.00 & 0.94 \\
                 &    &  &$U$& 0.0539 & 0.0350 & 0.76 0.41 & 0.34 & 0.13\\
NTTS~155913-2233 &7660&36&$Q$& 0.0597 & 0.0059 & 1.42 0.12 & 1.00 & 0.35 \\
                 &    &  &$U$& 0.0433 & 0.0059 & 1.09 0.12 & 0.78 & 0.01\\
                 &5550& 6&$Q$& 0.0604 & 0.0255 & 0.95 0.32 & 0.10 & 0.03\\
                 &    &  &$U$& 0.0697 & 0.0255 & 1.00 0.32 & 0.11 & 0.03\\
NTTS~160814-1857 &7660& 9&$Q$& 0.0956 & 0.0165 & 1.52 0.25 & 0.98 & 0.59\\
                 &    &  &$U$& 0.1040 & 0.0165 & 2.37 0.25 & 1.00 & 0.99\\
NTTS~160905-1859 &7660&34&$Q$& 0.0478 & 0.0070 & 1.15 0.12 & 0.89 & 0.03\\
                 &    &  &$U$& 0.0520 & 0.0070 & 1.32 0.12 & 0.99 & 0.18\\
                 &5550& 4&$Q$& 0.0867 & 0.0409 & 1.09 0.41 & 0.19 & 0.09\\
                 &    &  &$U$& 0.0423 & 0.0409 & 0.49 0.41 & 0.04 & 0.01\\
Haro~1-14C    &7660& 4&$Q$& 0.0104 & 0.0268 & 0.21 0.41 & 0.01 & 0.00 \\
              &    &  &$U$& 0.0831 & 0.0268 & 1.58 0.41 & 0.93 & 0.64\\
NTTS~162814-2427 &7660&22&$Q$& 0.1055 & 0.0097 & 1.91 0.15 & 1.00 & 0.96\\
                 &    &  &$U$& 0.2287 & 0.0097 & 3.34 0.15 & 1.00 & 1.00\\
NTTS~162819-2423S&7660&17&$Q$& 0.2430 & 0.0099 & 4.55 0.18 & 1.00 & 1.00\\
                 &    &  &$U$& 0.2527 & 0.0099 & 6.52 0.18 & 1.00 & 1.00\\
\enddata
\end{deluxetable}

\pagebreak \clearpage

\begin{deluxetable}{ll}
\tablewidth{0pt}
\tablecaption{Classification of the observed PMS binaries,
according to their variability at 7660\AA\label{Tab-Var}} 
\tablehead{
\colhead{Variability Classification} & {PMS Binary}}
\startdata
Variable & NTTS~160814-1857 (9),\\
         & NTTS~162819-2423S (17), NTTS~162814-2427 \tablenotemark{1} (22)\\  \\
Suspected variable & NTTS~155808-2219 (4), NTTS~155913-2233
\tablenotemark{1} (36) \\
         & NTTS~160905-1859 \tablenotemark{1} (34), Haro~1-14C (4)\\ \\
Possibly constant & \nodata \\ \\
Constant & \nodata
\tablenotetext{1}{These stars were sometimes observed to have very
different polarization and/or position angle values (well above or below
the majority of the data points) so their variability is based on data
excluding those atypical values.}
\tablecomments{The number of observations used for the variability tests
is indicated in parentheses.}
\enddata
\end{deluxetable}

\pagebreak \clearpage

\begin{deluxetable}{lcccccc}
\tablewidth{0pt}
\tablecaption{Polarization data for NTTS~155808-2219 at
7660\AA\label{Tab-n155808}} 
\tablehead{
\colhead{UT Date} & \colhead{JD} & \colhead{Phase\tablenotemark{1}} & 
\colhead{P} & \colhead{$\sigma(P)$} &
\colhead{$\theta$} & \colhead{$\sigma(\theta)$}\\
\colhead{} & 2400000.0$+$ & \colhead{} &
\colhead{(\%)} & \colhead{(\%)} &
\colhead{($\arcdeg$)} & \colhead{($\arcdeg$)}}
\startdata
1997 Apr 12 & 50550.849 & 0.454 & 0.481 & 0.063 &  152.7&    3.8\\
1997 Apr 16 & 50554.827 & 0.689 & 0.474 & 0.064 &  132.2&     3.8\\
1997 Jun 7  & 50606.708 & 0.755  & 0.522 & 0.102 &  133.5&    5.6\\
1998 May 1   & 50934.813 & 0.141  & 0.479 & 0.067 &  134.8&    4.0
\tablenotetext{1}{Calculated with the ephemeris $2449900.0+16.925E$
(Mathieu 1994).}
\enddata
\end{deluxetable}

\pagebreak \clearpage

\begin{deluxetable}{lcccccc}
\tablewidth{0pt}
\tablecaption{Polarization data for NTTS~155913-2233 at
7660\AA\label{Tab-n155913}}
\tablehead{
\colhead{UT Date} & \colhead{JD} & \colhead{Phase\tablenotemark{1}} & 
\colhead{P} & \colhead{$\sigma(P)$} &
\colhead{$\theta$} & \colhead{$\sigma(\theta)$}\\
\colhead{} & 2400000.0$+$ & \colhead{} &
\colhead{(\%)} & \colhead{(\%)} &
\colhead{($\arcdeg$)} & \colhead{($\arcdeg$)}}
\startdata
1995 May 2 &  49839.724   & 0.518  & 0.472  & 0.025   & 123.2    & 1.5 \\ 
1995 May 4 &  49841.771   & 0.362  & 0.456  & 0.037   & 122.4    & 2.3 \\
1995 May 7 &  49844.782   & 0.605  & 0.220  & 0.073   & 161.7    & 9.4 \\
1995 May 9 &  49846.667   & 0.382  & 0.579  & 0.037   & 122.1    & 1.8 \\
1995 May 10 &  49847.683   & 0.802  & 0.495  & 0.022   & 122.2    & 1.2 \\
1995 Aug 28 &  49957.543   & 0.128  & 0.500  & 0.040   & 120.3    & 2.3 \\
1995 Aug 31 &  49960.532   & 0.361  & 0.575  & 0.052   & 116.9    & 2.6   \\
1995 Sep 3 &  49963.531   & 0.598  & 0.424  & 0.066   & 123.1    & 4.5   \\
1996 Apr 19 &  50192.802   & 0.190  & 0.491  & 0.053   & 129.4    & 3.1   \\
1996 Apr 28 &  50201.789   & 0.898  & 0.443  & 0.030   & 126.9    & 1.9   \\
1996 Apr 29 &  50202.804   & 0.317  & 0.556  & 0.034   & 120.7    & 1.7   \\
1996 May 3 &  50206.780   & 0.957  & 0.482  & 0.040   & 123.4    & 2.4   \\
1996 May 7 &  50210.743   & 0.593  & 0.470  & 0.027   & 127.3    & 1.6   \\
1996 Jun 2 &  50236.698   & 0.301  & 0.467  & 0.069   & 123.9    & 4.3   \\
1996 Jul 7 &  50271.600   & 0.701  & 0.509  & 0.027   & 122.0    & 1.5   \\
1997 Feb 11 &  50490.909   & 0.183  & 0.486  & 0.028   & 125.0    & 1.6   \\
1997 Apr 2 &  50540.832   & 0.780  & 0.504  & 0.052   & 126.6    & 3.0   \\
1997 Apr 3 &  50541.811   & 0.184  & 0.471  & 0.033   & 121.3    & 2.0   \\
1997 Apr 5 &  50543.839   & 0.021  & 0.480  & 0.031   & 123.0    & 1.9   \\
1997 Apr 10 &  50548.786   & 0.062  & 0.461  & 0.029   & 129.1    & 1.8   \\
1997 Apr 11 &  50549.766   & 0.466  & 0.462  & 0.028   & 125.4    & 1.7   \\
1997 Apr 11 &  50549.872   & 0.510  & 0.476  & 0.046   & 168.2    & 2.8   \\
1997 Apr 12 &  50550.781   & 0.885  & 0.471  & 0.028   & 125.3    & 1.7   \\
1997 Apr 15 &  50553.786   & 0.125  & 0.472  & 0.027   & 121.5    & 1.6   \\
1997 Apr 16 &  50554.784   & 0.536  & 0.570  & 0.033   & 126.7    & 1.7   \\
1997 Jun 3 &  50602.627   & 0.276  & 0.474  & 0.045   & 123.9    & 2.7   \\
1997 Jun 4 &  50603.728   & 0.730  & 0.452  & 0.044   & 123.3    & 2.8   \\
1997 Jun 5 &  50604.678   & 0.122  & 0.429  & 0.034   & 127.3    & 2.3   \\
1997 Jun 6 &  50605.609   & 0.506  & 0.399  & 0.062   & 127.6    & 4.5   \\
1997 Jun 6 &  50605.725   & 0.554  & 0.584  & 0.074   & 120.8    & 3.6   \\
1997 Jun 7 &  50606.602   & 0.915  & 0.430  & 0.043   & 123.4    & 2.9   \\
1997 Jun 7 &  50606.738   & 0.972  & 0.420  & 0.052   & 126.2    & 3.5   \\
1997 Jun 9 &  50608.608   & 0.743  & 0.528  & 0.034   & 124.1    & 1.8   \\
1997 Jun 9 &  50608.741   & 0.798  & 0.536  & 0.043   & 127.4    & 2.3   \\
1997 Jun 15 &  50614.731   & 0.269  & 0.398  & 0.070   & 131.9    & 5.0   \\
1998 Apr 29 &  50932.838   & 0.514  & 0.420  & 0.054   & 127.4    & 3.7   \\
1998 May 1 &  50934.781   & 0.315  & 0.448  & 0.036   & 118.8    & 2.3   \\
1999 June 11 & 51340.609 & 0.751 & 0.567 & 0.034 & 124.6 & 1.7
\tablenotetext{1}{Calculated with the ephemeris $2445999.2+2.42378E$
(Mathieu et al.  1989).}  
\enddata
\end{deluxetable}

\pagebreak \clearpage

\begin{deluxetable}{lcccccc}
\tablewidth{0pt}
\tablecaption{Polarization data for NTTS~155913-2233 at
5550\AA\label{Tab-n155913-v}}
\tablehead{
\colhead{UT Date} & \colhead{JD} & \colhead{Phase\tablenotemark{1}} & 
\colhead{P} & \colhead{$\sigma(P)$} &
\colhead{$\theta$} & \colhead{$\sigma(\theta)$}\\
\colhead{} & 2400000.0$+$ & \colhead{} &
\colhead{(\%)} & \colhead{(\%)} &
\colhead{($\arcdeg$)} & \colhead{($\arcdeg$)}}
\startdata
1994 May 6 &  49479.557  & 0.924 & 0.518 & 0.048 & 128.4 & 2.3 \\ 
1994 May 8 &  49481.555  & 0.745 & 0.495 & 0.080 & 129.6 & 4.0\\
1994 May 9 &  49482.517  & 0.142 & 0.598 & 0.059 & 123.1 & 2.5\\
1994 May 10 &  49483.513 & 0.553 & 0.653 & 0.077 & 126.2 & 2.9\\
1994 May 11 &  49484.524 & 0.970 & 0.447 & 0.064 & 129.0 & 3.6\\
1994 May 12 &  49485.516 & 0.379 & 0.471 & 0.063 & 125.1 & 3.3
\tablenotetext{1}{Calculated with the ephemeris $2445999.2+2.42378E$
(Mathieu et al.  1989).}  
\enddata
\end{deluxetable}

\pagebreak \clearpage

\begin{deluxetable}{lcccccc}
\tablewidth{0pt}
\tablecaption{Polarization data for NTTS~160814-1857 at
7660\AA\ and 5550\AA\label{Tab-n160814}}
\tablehead{
\colhead{UT Date} & \colhead{JD} & \colhead{Phase\tablenotemark{1}} & 
\colhead{P} & \colhead{$\sigma(P)$} &
\colhead{$\theta$} & \colhead{$\sigma(\theta)$}\\
\colhead{} & 2400000.0$+$ & \colhead{} &
\colhead{(\%)} & \colhead{(\%)} &
\colhead{($\arcdeg$)} & \colhead{($\arcdeg$)}}
\startdata
1994 May 6\tablenotemark{2} & 49479.531 & 0.818 & 1.613 & 0.130 & 118.3 & 2.0\\
1995 May 9 & 49846.763 & 0.356 & 1.926 & 0.065 & 115.0 & 1.0\\
1996 May 8 & 50211.700 & 0.878 & 1.788 & 0.064 & 119.9 & 1.0\\
1997 Apr 11& 50549.737 & 0.214 & 1.909 & 0.048 & 116.3 & 1.0\\
1997 Apr 12& 50550.715 & 0.221 & 1.934 & 0.052 & 118.4 & 1.0\\
1997 Jun 5 & 50604.730 & 0.594 & 2.026 & 0.054 & 117.6 & 1.0\\
1998 Apr 29& 50932.660 & 0.860 & 1.817 & 0.035 & 116.0 & 1.0\\
1999 May 22& 51320.633 & 0.542 & 2.045 & 0.035 & 118.6 & 1.0\\
1999 Jun 11& 51340.733 & 0.680 & 2.079 & 0.076 & 118.0 & 1.0\\
1999 Jun 12& 51341.729 & 0.687 & 1.953 & 0.063 & 116.7 & 1.0
\tablenotetext{1}{Calculated with the ephemeris $2446003+ 144.7E$
(Mathieu et al.  1989).}
\tablenotetext{2}{Data obtained at 5550\AA.}
\enddata
\end{deluxetable}

\pagebreak \clearpage

\begin{deluxetable}{lcccccc}
\tablewidth{0pt}
\tablecaption{Polarization data for NTTS~160905-1859 at
7660\AA\label{Tab-n160905}}
\tablehead{
\colhead{UT Date} & \colhead{JD} & \colhead{Phase\tablenotemark{1}} & 
\colhead{P} & \colhead{$\sigma(P)$} &
\colhead{$\theta$} & \colhead{$\sigma(\theta)$}\\
\colhead{} & 2400000.0$+$ & \colhead{} &
\colhead{(\%)} & \colhead{(\%)} &
\colhead{($\arcdeg$)} & \colhead{($\arcdeg$)}}
\startdata
1995 May 9 & 49846.729 & 0.012  &  1.332 & 0.064  & 135.4  &  1.4\\
1996 May 3 & 50206.816 & 0.636  &  1.264 & 0.087  & 140.7  &  2.0\\
1996 May 7 & 50210.780 & 0.017  &  1.320 & 0.036  & 135.2  &  1.0\\
1996 Jun 2 & 50236.624 & 0.502  &  1.430 & 0.091  & 134.9  &  1.8\\
1997 Apr 2 & 50540.794 & 0.749  &  1.355 & 0.044  & 135.3  &  1.0\\
1997 Apr 3 & 50541.765 & 0.842  &  1.346 & 0.035  & 132.9  &  1.0\\
1997 Apr 5 & 50543.865 & 0.044  &  1.408 & 0.046  & 135.5  &  1.0\\
1997 Apr 10 & 50548.821 & 0.521  &  1.264 & 0.039  & 132.5  &  1.0\\
1997 Apr 11 & 50549.800 & 0.615  &  1.370 & 0.034  & 133.8  &  1.0\\
1997 Apr 12 & 50550.744 & 0.706  &  1.428 & 0.034  & 134.5  &  1.0\\
1997 Apr 15 & 50553.746 & 0.994  &  1.373 & 0.035  & 134.2  &  1.0\\
1997 Apr 16 & 50554.747 & 0.091  &  1.428 & 0.040  & 133.8  &  1.0\\
1997 Jun 3 & 50602.732 & 0.705  &  1.427 & 0.083  & 135.4  &  1.7  \\
1997 Jun 4 & 50603.616 & 0.790  &  1.389 & 0.053  & 135.2  &  1.1\\
1997 Jun 5 & 50604.595 & 0.884  &  1.395 & 0.050  & 132.3  &  1.0\\
1997 Jun 6 & 50605.693 & 0.989  &  1.307 & 0.055  & 135.5  &  1.2\\
1997 Jun 7 & 50606.678 & 0.084  &  1.332 & 0.045  & 133.7  &  1.0\\
1997 Jun 9 & 50608.675 & 0.276  &  1.486 & 0.035  & 134.2  &  1.0 \\
1997 Jun 15 & 50614.687 & 0.854  &  1.370 & 0.043  & 134.4  &  1.0\\
1997 Jul 7 & 50636.626 & 0.964  &  1.353 & 0.048  & 134.4  &  1.0\\
1997 Jul 11 & 50640.605 & 0.346  &  1.443 & 0.041  & 134.8  &  1.0\\
1997 Sep 9 & 50700.541 & 0.109  &  1.426 & 0.065  & 132.4  &  1.3\\
1998 Apr 27 & 50930.708 & 0.241  &  1.316 & 0.034  & 133.2  &  1.0\\
1998 Apr 29 & 50932.713 & 0.433  &  1.315 & 0.032  & 132.8  &  1.0\\
1998 Apr 30 & 50933.701 & 0.528  &  1.318 & 0.031  & 133.0  &  1.0\\
1998 May 1 & 50934.670 & 0.622  &  1.419 & 0.036  & 120.7  &  1.0\\
1998 May 13 & 50946.745 & 0.783  &  1.333 & 0.052  & 134.8  &  1.1\\
1998 May 14 & 50947.768 & 0.881  &  1.409 & 0.027  & 135.0  &  1.0\\
1998 May 20 & 50953.720 & 0.453  &  1.435 & 0.030  & 135.7  &  1.0\\
1998 May 25 & 50958.613 & 0.924  &  1.418 & 0.033  & 135.0  &  1.0\\
1998 May 27 & 50960.627 & 0.118  &  1.427 & 0.051  & 134.3  &  1.0\\
1998 Jun 1 & 50966.634 & 0.695  &  1.448 & 0.052  & 133.6  &  1.0\\
1999 May 22 & 51320.698 & 0.740 & 1.417 & 0.043 & 135.5 &  1.0\\
1999 Jun 12 &  51341.610 & 0.750 & 1.400 & 0.048 & 134.7 & 1.0\\
1999 Jun 13 &  51342.611 & 0.847 & 1.413 & 0.050 & 133.9 & 1.0\\
1999 Jun 14 &  51343.601 & 0.942 & 1.411 & 0.055 & 133.5 & 1.1
\tablenotetext{1}{Calculated with the ephemeris $2445998.6+10.40E$
(Mathieu et al.  1989).}
\enddata
\end{deluxetable}

\pagebreak \clearpage

\begin{deluxetable}{lcccccc}
\tablewidth{0pt}
\tablecaption{Polarization data for NTTS~160905-1859 at
5550\AA\label{Tab-n160905-v}}
\tablehead{
\colhead{UT Date} & \colhead{JD} & \colhead{Phase\tablenotemark{1}} & 
\colhead{P} & \colhead{$\sigma(P)$} &
\colhead{$\theta$} & \colhead{$\sigma(\theta)$}\\
\colhead{} & 2400000.0$+$ & \colhead{} &
\colhead{(\%)} & \colhead{(\%)} &
\colhead{($\arcdeg$)} & \colhead{($\arcdeg$)}}
\startdata
1994 May 6  & 49479.510 & 0.702 & 1.059 & 0.087 & 132.6 & 2.0\\
1994 May 8  & 49481.597 & 0.903 & 1.068 & 0.105 & 136.3 & 2.5\\
1994 May 9  & 49482.549 & 0.995 & 1.082 & 0.0065 & 132.0 & 1.5\\
1994 May 11 & 49484.583 & 0.190 & 1.150 & 0.0085 & 136.2 & 2.0
\tablenotetext{1}{Calculated with the ephemeris $2445998.6+10.40E$
(Mathieu et al.  1989).}
\enddata
\end{deluxetable}

\pagebreak \clearpage

\begin{deluxetable}{lcccccc}
\tablewidth{0pt}
\tablecaption{Polarization data for Haro 1-14C at
7660\AA\label{Tab-haro}}
\tablehead{
\colhead{UT Date} & \colhead{JD} & \colhead{Phase\tablenotemark{1}} & 
\colhead{P} & \colhead{$\sigma(P)$} &
\colhead{$\theta$} & \colhead{$\sigma(\theta)$}\\
\colhead{} & 2400000.0$+$ & \colhead{} &
\colhead{(\%)} & \colhead{(\%)} &
\colhead{($\arcdeg$)} & \colhead{($\arcdeg$)}}
\startdata
1997 Apr 11 & 50549.860 & 0.099  & 0.989 & 0.055 &   31.3 &   1.6\\
1997 Jun 5 & 50604.749 & 0.192  & 1.074 & 0.199 &  133.2 &   5.3\\
1997 Jun 6 & 45605.750 & 0.733  & 1.030 & 0.096 &   31.3 &   2.7\\
1997 Jul 11 & 50640.637 & 0.253  & 1.085 & 0.043 &   31.9 &   1.1\\
1998 May 1 & 50934.760 & 0.750  & 1.166 & 0.049 &   33.2 &   1.2
\tablenotetext{1}{Calculated with the ephemeris $49900.0+ 591.0E$
(period from Mathieu 1994).}
\enddata
\end{deluxetable}

\pagebreak \clearpage

\begin{deluxetable}{lcccccc}
\tablewidth{0pt}
\tablecaption{Polarization data for NTTS~162814-2427 at
7660\AA\label{Tab-n162814}}
\tablehead{
\colhead{UT Date} & \colhead{JD} & \colhead{Phase\tablenotemark{1}} & 
\colhead{P} & \colhead{$\sigma(P)$} &
\colhead{$\theta$} & \colhead{$\sigma(\theta)$}\\
\colhead{} & 2400000.0$+$ & \colhead{} &
\colhead{(\%)} & \colhead{(\%)} &
\colhead{($\arcdeg$)} & \colhead{($\arcdeg$)}}
\startdata
1995 May 10 &  49847.723  & 0.206 & 3.371 & 0.080  &  20.9 &   1.0\\
1996 May 7 &  50210.808  & 0.306 & 3.261 & 0.049  &  22.5 &   1.0\\
1996 Jul 9 &  50273.616  & 0.053 & 3.210 & 0.062  &  21.7 &   1.0\\
1997 Apr 3 &  50541.841  & 0.514 & 3.414 & 0.050  &  23.4 &   1.0\\
1997 Apr 10 &  50548.845  & 0.709 & 3.556 & 0.046  &  23.1 &   1.0\\
1997 Apr 11  &  50549.828  & 0.736 & 3.444 & 0.047  &  23.0 &   1.0\\
1997 Apr 12  &  50550.813  & 0.763 & 3.417 & 0.044  &  22.7 &   1.0\\
1997 Apr 15  &  50553.824  & 0.847 & 3.415 & 0.044  &  21.9 &   1.0\\
1997 Jun 3 &  50602.677  & 0.206 & 2.626 & 0.033  &  15.0 &   1.0\\
1997 Jun 4 &  50603.679  & 0.234 & 3.413 & 0.042  &  22.9 &   1.0\\
1997 Jun 5 &  50604.635  & 0.261 & 3.550 & 0.040  &  22.6 &   1.0\\
1997 Jun 6 &  50605.650  & 0.289 & 3.459 & 0.049  &  23.2 &   1.0\\
1997 Jun 7 &  50606.638  & 0.316 & 3.346 & 0.054  &  22.9 &   1.0\\
1997 Jun 9 &  50608.640  & 0.372 & 3.694 & 0.038  &  24.1 &   1.0\\
1997 Jun 15 &  50614.629  & 0.539 & 3.667 & 0.070  &  22.5 &   1.0\\
1997 Sep 9 &  50700.523  & 0.928 & 3.595 & 0.063  &  20.6 &   1.0\\
1998 Apr 27 &  50930.754  & 0.332 & 3.557 & 0.038  &  22.7 &   1.0\\
1998 Apr 29 &  50932.753  & 0.388 & 3.445 & 0.036  &  22.9 &   1.0\\
1998 Apr 30 &  50933.748  & 0.415 & 3.621 & 0.034  &  22.7 &   1.0\\
1998 May 1 &  50934.721  & 0.442 & 3.674 & 0.032  &  23.0 &   1.0\\
1998 May 25 &  50958.643  & 0.108 & 3.501 & 0.037  &  22.2 &   1.0\\
1999 Jun 13 & 51342.650 & 0.789 & 3.489 & 0.063 & 22.6 & 1.0\\
1999 Jun 14 & 51343.632 & 0.817 & 3.085 & 0.150 & 15.3 & 1.4
\tablenotetext{1}{Calculated with the ephemeris $2445023.0+35.95E$
(Mathieu et al.  1989).}
\enddata
\end{deluxetable}

\pagebreak \clearpage

\begin{deluxetable}{lcccccc}
\tablewidth{0pt}
\tablecaption{Polarization data for NTTS~162819-2423S at
7660\AA\label{Tab-n162819}}
\tablehead{
\colhead{UT Date} & \colhead{JD} & \colhead{Phase\tablenotemark{1}} & 
\colhead{P} & \colhead{$\sigma(P)$} &
\colhead{$\theta$} & \colhead{$\sigma(\theta)$}\\
\colhead{} & 2400000.0$+$ & \colhead{} &
\colhead{(\%)} & \colhead{(\%)} &
\colhead{($\arcdeg$)} & \colhead{($\arcdeg$)}}
\startdata
1995 May 7 & 49844.838  & 0.196  &  3.639 &  0.141  &   14.1  &   1.1\\
1996 May 7 & 50210.828  & 0.304  &  3.041 &  0.083  &   18.2  &   1.0\\
1996 May 8 & 50211.768  & 0.315  &  2.528 &  0.052  &   13.6  &   1.0\\
1996 Jun 1 & 50235.667  & 0.583  &  2.637 &  0.061  &   15.7  &   1.0\\
1996 Jul 7 & 50271.619  & 0.986  &  2.829 &  0.028  &   14.2  &   1.0\\
1997 Apr 3 & 50541.865  & 0.019  &  2.651 &  0.029  &   14.7  &   1.0\\
1997 Apr 10 & 50548.868  & 0.098  &  2.871 &  0.037  &   17.6  &   1.0\\
1997 Apr 15 & 50553.865  & 0.154  &  2.940 &  0.047  &   14.6  &   1.0\\
1997 Jun 3 & 50602.677 &  0.702  &  2.626  &  0.033 &  15.0 & 1.0 \\
1997 Jun 5 & 50604.706  & 0.725  &  2.887 &  0.036  &   15.0  &   1.0\\
1997 Jun 9 & 50608.708  & 0.770  &  3.549 &  0.034  &   18.2  &   1.0\\
1997 Jun 9 & 50608.714  & 0.770  &  3.208 &  0.048  &   17.1  &   1.0\\
1998 Apr 29 & 50932.808  & 0.407  &  2.995 &  0.029  &   17.6  &   1.0\\
1999 Jun 11 & 51340.651 & 0.984 & 2.842 & 0.044 & 15.7 &  1.0\\
1999 Jun 12 & 51341.683 & 0.996 & 3.218 & 0.052 & 16.5 &  1.0\\
1999 Jun 13 & 51342.695 & 0.007 & 3.190 & 0.053 & 18.0 &  1.0\\ 
1999 Jun 13 & 51342.736 & 0.008 & 2.977 & 0.065 & 16.9 &  1.0 
\tablenotetext{1}{Calculated with the ephemeris $2445996.0+ 89.1E$
(Mathieu et al.  1989).}
\enddata
\end{deluxetable}

\pagebreak \clearpage

\begin{deluxetable}{lcccccccc}
\tablewidth{0pt}
\tablecaption{Noise analysis and orbital inclination from the BME model
for some observed binaries\label{Tab-noiseBME}} 
\tablehead{
\colhead{Star} & \colhead{$DQ$} & \colhead{$\gamma$} & 
\colhead{Noise\tablenotemark{1}} & \colhead{Noise} & 
\colhead{$i({\cal O}2)$} & \colhead{$\sigma(i({\cal O}2)$)} & 
\colhead{$i({\cal O}1)$} & \colhead{$\sigma(i({\cal O}1)$)}\\
\colhead{} & \colhead{} & \colhead{} & 
\colhead{for $Q$} & \colhead{for $U$} & 
\colhead{(\arcdeg)} & \colhead{(\arcdeg)} &
\colhead{(\arcdeg)} & \colhead{(\arcdeg)}}
\startdata
NTTS~155913-2233 & 0.202 & 12.2 & 0.16 & 0.22 & 96.7 & 8.8 & 23.4 & 84.5\\
..... set 1      & 0.276 & 6.6  & 0.23 & 0.24 & 95.5 & 9.7 & 87.8 & 17.3\\
..... set 2      & 0.248 & 8.1 & 0.20 & 0.23 & 103.0 & 19.4 & 88.6 & 18.7\\
NTTS~160814-1857 & 0.941 & 45.8 & 0.23 & 0.23 & 97.1 & 0.3 & 93.4 & 0.6\\
NTTS~160905-1859 & 0.281 & 6.3  & 0.29 & 0.24 & 84.4 & 8.1 & 150.2 & 68.8\\
..... set 1      & 0.106 & 44.6 & 0.34 & 0.13 & 95.6 & 7.7 & 100.0 & 15.8\\
..... set 2      & 0.190 & 13.9 & 0.34 & 0.15 & 93.2 & 14.4 & 66.8 & 37.4\\
..... set 3      & 0.143 & 24.3 & 0.10 & 0.31 & 98.4 & 17.1 & 99.4 & 40.7\\
..... set 4      & 0.256 & 7.6  & 0.44 & 0.51 & 112.2 & 35.0 & 86.6 & 25.1\\
NTTS~162814-2427\tablenotemark{2} & 0.099&51.1&0.22&0.10 &
	\nodata&\nodata& 93.8 & 2.6
\tablenotetext{1}{The noise is the square root of the variance of the fit
over the amplitude of the variations; the amplitude comes from the
maximum values of the data and not from the fit.}
\tablenotetext{2}{Since NTTS~162814-2427's eccentricity is above 0.3,
only the BME results obtained from the first order make sense (Paper~I).}
\enddata
\end{deluxetable}

\pagebreak \clearpage

\begin{deluxetable}{lccccc}
\tablewidth{0pt}
\tablecaption{Other parameters returned by the BME model: $\Omega$, the
orientation of the orbital plane, and moments of the distribution of the
scatterers. \label{Tab-omegagammas}}
\tablehead{
\colhead{Star} & \colhead{$\Omega$} & \colhead{$\sigma(\Omega)$} &
\colhead{$\tau_0 G$} & \colhead{$\tau_0 H$} & 
\colhead{$\tau_0 H / \tau_0 G$}\\
\colhead{} & \colhead{($\arcdeg$)} & \colhead{($\arcdeg$)} & 
\colhead{$\times 10^{-4}$} & 
\colhead{$\times 10^{-4}$} & \colhead{}}
\startdata
NTTS~155913-2233 & 14.1  & 17.6  & 2.5  & 2.9  & 1.3 \\
NTTS~160814-1857 & 133.8 & 0.6   & 7.5  & 16   & 2.3 \\
NTTS~160905-1859 & 95.8  & 20.7  & 1.6  & 2.8  & 1.8 \\
NTTS~162814-2427 & 0.6   & 3.0   & 7.4  & 6.1  & 0.8 \\
NTTS~162819-2423S& 46.0  & 0.3   & 9.3  & 31   & 3.3 \\
\enddata           
\end{deluxetable}

\pagebreak \clearpage

\begin{deluxetable}{lcccccc}
\tablecolumns{7}
\tablewidth{0pc}
\tablecaption{Ratio of the amplitude of the single-periodic variations
over the double-periodic ones for the polarization.\label{Tab-lambda}} 
\tablehead{
\colhead{} & \multicolumn{4}{c}{Ratio of the amplitude in $1\lambda$} & 
\colhead{} & \colhead{}\\
\colhead{} & \multicolumn{4}{c}{over the amplitude in $2\lambda$} &
\colhead{} & \colhead{}\\
\cline{2-5}\\
\colhead{Star} & \colhead{$Q$} & \colhead{$U$} & \colhead{$P$} &
\colhead{$\theta$} & \colhead{$e$} & \colhead{Period} \\
\colhead{} & \colhead{} & \colhead{} & \colhead{} & \colhead{} & 
\colhead{} & \colhead{(d)}}
\startdata 
LkCa~3     & 1.62 & 0.32 & 2.98 & 0.71          & 0.20 & 12.9\\
V826~Tau   & 3.31 & 0.49 & 0.63 & 0.75          & 0.0  & 3.9\\
GW~Ori     & 0.50 & 1.52 & 0.93 & 0.67          & 0.04 & 242\\
Par~1540   & 0.57 & 0.45 & 0.44 & 0.50          & 0.12 & 33.7\\
Par~2494   & 0.56 & 0.68 & 0.69 & 0.56          & 0.26 & 19.5\\
W~134      & 0.60 & 0.60 & 0.68 & 0.29          & 0    & 6.4\\
NTTS~155913-2233  & 0.68 &  1.86 & 1.02 & 0.78  & 0.02 & 2.4\\
NTTS~160814-1857  & 1.26 &  0.13 & 1.80 & 0.72  & 0.26 & 145\\
NTTS~160905-1859  & 2.43 &  0.49 & 0.49 & 2.29  & 0.17 & 10.4\\
..... set1 &  1.46 &  1.05 & 1.03 & 1.43        & 0.17 & 10.4\\
..... set2 &  1.98 &  0.58 & 0.58 & 1.95        & 0.17 & 10.4\\
..... set3 &  0.91 &  0.59 & 0.59 & 0.91        & 0.17 & 10.4\\
..... set4 &  0.79 &  1.49 & 1.49 & 0.82        & 0.17 & 10.4\\
NTTS~162814-2427&0.52 & 8.11 & 1.74 & 2.73      & 0.48 & 36\\
NTTS~162819-2423S&0.79 &  0.36 & 0.62 & 0.31    & 0.41 & 89\\
AK~Sco\tablenotemark{1} & 7.34 &  1.14 & 1.09 & 4.15           & 0.47 & 13.6\\
EK~Cep\tablenotemark{1} & 0.75 &  1.55 & 0.61 & 4.07           & 0.11 & 4.4
\tablecomments{The orbital eccentricities and period given are
approximate values only.}
\tablenotetext{1}{Data for AK~Sco and EK~Cep will be presented in future
papers.}
\enddata
\end{deluxetable}

\end{document}